\documentclass[arguments]{aastex63}

\submitjournal{The Astrophysical Journal}

\shorttitle{Component spectral Variation}
\shortauthors{Basu, Mitra \& Melikidze}

\begin{document}

\title{Spectral variation across Pulsar Profile due to Coherent Curvature Radiation}


\author[0000-0003-1824-4487]{Rahul Basu}
\affiliation{Janusz Gil Institute of Astronomy, University of Zielona G\'ora, ul. Szafrana 2, 65-516 Zielona G\'ora, Poland.}
\affiliation{Inter-University Centre for Astronomy and Astrophysics, Pune, 411007, India.}

\author[0000-0002-9142-9835]{Dipanjan Mitra}
\affiliation{National Centre for Radio Astrophysics, Tata Institute of Fundamental Research, Pune 411007, India.}
\affiliation{Janusz Gil Institute of Astronomy, University of Zielona G\'ora, ul. Szafrana 2, 65-516 Zielona G\'ora, Poland.}

\author[0000-0003-1879-1659]{George I. Melikidze}
\affiliation{Janusz Gil Institute of Astronomy, University of Zielona G\'ora, ul. Szafrana 2, 65-516 Zielona G\'ora, Poland.}
\affiliation{Evgeni Kharadze Georgian National Astrophysical Observatory, 0301 Abastumani, Georgia.}

\begin{abstract}
The pulsar profile is characterised by two distinct emission components, the 
core and the cone. The standard model of a pulsar radio emission beam 
originating from dipolar magnetic fields, places the core at the centre 
surrounded by concentric layers of inner and outer conal components. The core 
emission is expected to have steeper spectra compared to the cones. We present 
a detailed analysis of the relative differences in spectra between the core and 
conal emission from a large sample of 53 pulsars over a wide frequency range 
between 100 MHz and 10 GHz. The core was seen to be much steeper than the cones
particularly between 100 MHz and 1 GHz with a relative difference between the 
spectral index $\Delta\alpha_{core/cone}\sim$ -1.0. In addition we also 
found the spectra of the outer conal components to be steeper than the inner 
cone with relative difference in the spectral index $\Delta\alpha_{in/out}\sim$
+0.5. The flattening of the spectra from the magnetic axis towards the edge of 
the open field line region with increasing curvature of the field lines is a 
natural outcome of the coherent curvature radiation from charged soliton 
bunches, and explains the difference in spectra between the core and the cones. 
In addition due to the relativistic beaming effect, the radiation is only 
visible when it is directed towards the observer over a narrow angle 
$\theta\leq1/\gamma$, where $\gamma$ is the Lorentz factor of the outflowing 
plasma clouds. This restricts the emission particularly from outer cones, that 
are associated with field lines with larger curvature thereby making the 
spectra steeper than the inner cones.
\end{abstract}


\section{Introduction}
\noindent
The radio emission from normal period pulsars ($P >$ 0.1 second) with measured 
flux density over a wide frequency range, between 100 MHz and 10 GHz, exhibit a 
inherently steep power law spectra with typical spectral index, $\alpha\sim
-1.8$ \citep[$S\propto\nu^{\alpha}$,][]{LYL95,MKK00,JvK18}. In most cases the 
spectral index is obtained from the total flux density estimated across the 
average pulse profile. The pulsar profiles have complex shapes, usually 
comprising of several Gaussian like components, and detailed phenomenological 
studies suggest that the components have specific locations within the pulse 
window. The pulsar radio emission beam has a roughly circular shape, and can be
described by the so called `core-cone' model. According to this model the beam 
consists of a central core surrounded by two rings of nested conal emission, 
namely the inner and the outer cones \citep{ET_R90,ET_R93,MD99}. The width and
shape of the pulse profile depends on the pulsar geometry and observers' line 
of sight (LOS) traverse across the emission beam. Central LOS cuts form 
core-cone Triple (T) and core-double cone multiple (M) profile types, while the
profile shape becomes conal Quadruple ($_c$Q), conal Triple ($_c$T) and double 
(D) shaped as the LOS traverses are progressively away from the magnetic axis.
T$_{1/2}$ is another form of core-cone profile with two components where a 
conal component on one side of the core is either missing or too weak to detect
at certain frequencies. A core single (S$_t$) profile signifies a central LOS 
traverse where the conal components are absent while a conal single (S$_d$) 
profile corresponds to peripheral LOS cuts. Table \ref{tab:ProfClass} 
summarises the classification scheme of the pulsar profile types. It has also 
been shown that the distribution of both the core and the conal component 
widths with period has similar lower boundary lines which is proportional to 
$P^{-0.5}$ \citep{MGM12,SBM18}, suggesting similar emission heights for both. 
The pulsar emission is usually highly polarized and shows a polarization 
position angle (PPA) swing across the profile that resemble a S-shape curve. 
This is a result of the emission originating from strictly dipolar magnetic 
field region \citep[Rotating Vector Model,][]{RC69}. One of the most 
significant details about the origin of radio emission comes from constrains on
its location at heights of 100-1000 km from the stellar surface. These are 
obtained from two independent estimates, the geometrical morphology described 
above \citep{ET_R93,KG98,ET_MR02,KG03} and the abberation-retardation effect 
that shifts the PPA from the profile center, with the displacement being
proportional to the emission height \citep{BCW91,P92,XKJ96,ML04,WJ08,KMG09}. 

\begin{deluxetable}{cccl}
\tablenum{1}
\tabletypesize{\normalsize}
\tablecaption{Summary of Profile Classification\label{tab:ProfClass}}
\tablewidth{0pt}
\tablehead{
 \colhead{Symbol} & \colhead{Type} & \colhead{No. of Components} & \colhead{LOS Cut of Emission Beam}}
\startdata
  S$_d$ & Cone & 1 & Outer Edge \\
  S$_t$ & Core & 1 & Center \\
  D & Cone & 2 & Closer to the outer Edge \\
  T$_{1/2}$ & Core-Cone & 2 & Center \\
  $_c$T & Cone & 3 & Middle \\
  T & Core-Cone & 3 & Center \\
  $_c$Q & Cone & 4 & Closer to the Center \\
  M & Core-Cone & 5 & Center \\
\enddata
\end{deluxetable}

Once it is established that the emission arises from a narrow range of heights 
in the dipolar magnetic field region, a gradual variation in certain emission 
properties is also expected due to change in curvature, from the central field 
lines near the axis towards the edge of the open field line region. This is 
reflected in the PPA particularly in the central LOS traverses where the core 
component is associated with large PPA variations, while the PPA traverse is 
relatively flat across the cones \citep{ET_R90,ET_R93}. Apart from having 
specific location within the pulsar beam, the core and conal flux densities 
exhibit different spectra, with the core emission having a steeper spectra 
compared to the cones \citep{ET_R83}. This effect is once again best observed 
in T and M profiles, when the core, inner and outer cones are clearly visible 
across different frequencies. However, it is more complicated to measure the 
flux densities from pulsars as they require detailed instrumental calibration 
for scaling the flux level of the measured signal. The pulsar emission also 
shows variability due to scintillation in the intervening medium \citep{R90}, 
and hence it is essential to average over multiple observing sessions to find a
mean level. A way around the issues with flux measurement has been suggested in
the recent work of \cite{BMM21}, where it was noted that the different 
components in the profile are equally affected by the flux scaling and 
scintillation issues and hence the ratio between their intensities is 
unaffected by these variations. The frequency evolution of the ratio between 
the components can be used to find the difference in spectra between the core 
and the different types of conal emission. The spectral difference between the 
core and the conal emission in 21 pulsars observed in the MSPES survey at 325 
and 610 MHz \citep{MBM16} were estimated and the core was found to be steeper 
than the cone with $\Delta\alpha_{core/cone}\sim-0.7$. In addition it was also 
found that the inner cone was less steep compared to the outer cones where
$\Delta\alpha_{in/out}\sim+0.3$. 

Currently there are no physical models that can explain why the pulsar spectra 
varies across the emission beam? On the other hand several observational 
features in pulsars like the radio emission heights, polarization properties, 
etc., suggest that the pulsar emission can be excited by curvature radiation 
from charge bunches in a relativistically flowing electron-positron plasma 
\citep{RS75,MGP00,GLM04,MGM09,MMG14,M17}. These charge bunches are formed due 
to nonlinear growth of plasma instability that leads to formation of 
relativistic, charged solitons \citep{GLM04,MMG14,LMM18,RMM20}. \cite{MGP00} 
also showed that there are significant differences between the spectrum of a 
single particle curvature radiation and a charge bunch of finite length. Thus 
using the constraints from observations and the theory of coherent curvature 
radiation it is now possible to address the origin of pulsar spectral index and
its variation across different components.

In this paper we aim to carry out an exhaustive study of the spectra of the 
pulsar profile. A large number of average profiles in a wide frequency range 
has been measured and made publicly available over many decades of pulsar 
studies \citep{LYL95,SGG95,vHX97,KKW98,GL98,WCL99,BKK16,MBM16,JK18}. We have 
estimated the spectral variations of the different component types in 53 
pulsars, from archival profiles over a wide frequency range, between 100 MHz 
and 10 GHz. We also explore the variations expected in the spectra across the 
emission beam due to curvature radiation from charge bunches and compare with 
the measurements.

\section{Average Profile analysis}\label{sec:obs}

The primary analysis concerns with estimating the relative differences in 
spectra between different component types, core, inner cone and outer cones, 
within the pulsar average profile. Following the suggestion of \cite{BMM21}, 
the spectral difference between between different component types is estimated 
as :
\begin{eqnarray}
S_{core} & = & S_1~ \nu^{\alpha_{core}} \nonumber\\
S_{cone} & = & S_2~ \nu^{\alpha_{cone}} \nonumber\\
S_{in} & = & S_3~ \nu^{\alpha_{in}} \nonumber\\
S_{out} & = & S_4~ \nu^{\alpha_{out}} \nonumber\\
\left(\frac{S_{core}}{S_{cone}}\right) & = & \left(\frac{S_1}{S_2}\right) \nu^{\Delta\alpha_{core/cone}} ~~~~~ \Delta\alpha_{core/cone} = \alpha_{core} - \alpha_{cone} \nonumber \\
\left(\frac{S_{core}}{S_{in}}\right) & = & \left(\frac{S_1}{S_3}\right) \nu^{\Delta\alpha_{core/in}} ~~~~~ \Delta\alpha_{core/in} = \alpha_{core} - \alpha_{in} \nonumber \\
\left(\frac{S_{core}}{S_{out}}\right) & = & \left(\frac{S_1}{S_4}\right) \nu^{\Delta\alpha_{core/out}} ~~~~~ \Delta\alpha_{core/out} = \alpha_{core} - \alpha_{out} \nonumber \\
\left(\frac{S_{in}}{S_{out}}\right) & = & \left(\frac{S_3}{S_4}\right) \nu^{\Delta\alpha_{in/out}} ~~~~~ \Delta\alpha_{in/out} = \alpha_{in} - \alpha_{out} \nonumber \\
\end{eqnarray}

\begin{deluxetable}{ccccccccc}
\tablenum{2}
\tabletypesize{\footnotesize}
\tablecaption{Pulsar List\label{tab:srclist}}
\tablewidth{0pt}
\tablehead{ 
    & \colhead{PSR} & \colhead{Period} & \colhead{Class} & \colhead{Spectral Index} & \colhead{Incl. angle} & \colhead{Freq. Range} & \colhead{Nfreq} & \colhead{Method} \\
   &   & \colhead{(s)} &   &   & ($\degr$) &  \colhead{(MHz)} &   &   }
\startdata
  1 &  B0203-40  & 0.631 & T$_{1/2}$ & --- & 72 & 325 -- 1400 & 2  & Peak \\
  2 &  B0329+54  & 0.715 & T & -1.6 & 30 & 140 -- 4850 & 9 & Total \\
  3 &  B0450+55  & 0.341 & T & -1.2 & 32 & 325 -- 4850 & 7 & Avg \\
  4 &  B0621-04  & 1.039 & M & -1.0 & 32 & 410 -- 1408 & 3 & Peak \\
  5 &  B0626+24  & 0.477 & T$_{1/2}$ & -1.5 & 30 & 170 -- 4850 & 4 & Peak \\
  6 &  B0844-35  & 1.116 & $_c$Q & -2.0 & 26 & 325 -- 610 & 2 & Avg \\
  7 &  B0919+06  & 0.431 & T & -1.8 & 48 & 135 -- 610 &  3 & Avg \\
  8 &  B0940-55  & 0.664 & T & -2.6 & 34 & 1400 -- 3100 & 2 & Total \\
  9 & J1034-3224 & 1.151 & $_c$Q & -1.6 & 20 & 325 -- 1400 & 4 & Total \\
 10 &  B1046-58  & 0.124 & $_c$T & -0.5 & --- & 1400 -- 8356 & 2 & Total \\
 11 & J1141-3322 & 0.291 & T & -1.2 & 36 & 436 -- 1400 & 2 & Total \\
 12 &  B1154-62  & 0.401 & T & -2.4 & 17 & 1400 -- 3100 & 2 & Total \\
 13 &  B1237+25  & 1.382 & M & -1.8 & 53 & 120 -- 1400 & 5 & Avg \\
 14 &  B1323-58  & 0.478 & T$_{1/2}$ & -1.8 & 52 & 1400 -- 8356 & 3 & Peak \\
 15 &  B1325-49  & 1.479 & M & --- & $\sim$90 & 325 - 610 & 2 & Total \\
 16 &  B1353-62  & 0.456 & T & -1.8 & 27 & 1400 -- 3100 & 2 & Total \\
 17 &  B1508+55  & 0.740 & T & -2.3 & 45 & 140 -- 325 & 3 & Total \\
 18 &  B1541+09  & 0.748 & T & -2.1 & 5 & 140 -- 1418 & 5 & Avg \\
 19 & J1557-4258 & 0.329 & T & --- & 67 & 610 -- 1400 & 2 & Total \\
 20 &  B1556-44  & 0.257 & T & -2.3 & 32 & 325 -- 1560 & 3 & Peak \\
 21 &  B1600-49  & 0.327 & T & -1.6 & $\sim$90 & 610 -- 1400 & 2 & Total \\
 22 & J1625-4048 & 2.355 & T & --- & 29 & 436 -- 1400 & 3 & Total \\
 23 &  B1642-03  & 0.388 & T & -2.3 & 70 & 610 -- 10550 & 5 & Avg \\
 24 &  B1700-32  & 1.212 & T & -1.5 & 48 & 325 -- 610 & 2 & Total \\
 25 &  B1732-07  & 0.419 & T & -1.8 & 64 & 325 -- 1400 & 3 & Avg \\
 26 &  B1737+13  & 0.803 & M & -1.4 & 41 & 325 -- 1418 & 5 & Total \\
 27 &  B1738-08  & 2.043 & $_c$Q & -2.5 & 26 & 325 -- 610 & 2 & Avg \\
 28 &  B1758-29  & 1.082 & T & -2.0 & 36 & 325 -- 1400 & 3 & Total \\
 29 &  B1804-08  & 0.164 & T & -1.4 & 63 & 610 -- 3100 & 4 & Total \\
 30 &  B1821+05  & 0.753 & T & -1.7 & 32 & 325 -- 4850 & 7 & Total \\
 31 &  B1826-17  & 0.307 & T & -1.7 & 39 & 925 -- 1408 & 2 & Total \\
 32 &  B1831-03  & 0.687 & T & -2.8 & 54 & 925 -- 1400 & 2 & Peak \\
 33 &  B1831-04  & 0.290 & M & -1.3 & 10 & 325 -- 610 & 2 & Avg \\
 34 &  B1839+09  & 0.381 & T & -2.0 & 83 & 130 -- 4850 & 4 & Peak \\
 35 &  B1857-26  & 0.612 & M & -1.2 & 25 & 325 -- 1400 & 5 & Total \\
 36 &  B1859+03  & 0.655 & T$_{1/2}$ & -2.7 & 35 & 925 -- 1642 & 3 & Peak \\
 37 &  B1907+00  & 1.017 & T & -1.8 & 69 & 610 -- 1418 & 2 & Total \\
 38 &  B1907+10  & 0.284 & T$_{1/2}$ & -1.9 & 49 & 610 -- 1400 & 2 & Peak \\
 39 &  B1907-03  & 0.505 & T & -2.6 & 33 & 610 -- 1420 & 2 & Avg \\
 40 &  B1911+13  & 0.521 & T & -1.4 & 52 & 606 -- 1418 & 2 & Peak \\
 41 &  B1914+09  & 0.270 & T$_{1/2}$ & -2.3 & 52 & 325 -- 1642 & 6 & Total \\
 42 &  B1917+00  & 1.272 & T & -2.3 & 81 & 325 -- 1642 & 5 & Total \\
 43 &  B1918+26  & 0.786 & T$_{1/2}$ & -1.3 & 44 & 170 -- 1418 & 2 & Peak \\
 44 &  B1920+21  & 1.078 & T$_{1/2}$ & -2.4 & 44 & 610 -- 1642 & 4 & Peak \\
 45 &  B1929+10m & 0.227 & T & -1.7 & $\sim$90 & 120 -- 10550 & 10 & Peak \\
 46 &  B1946+35  & 0.717 & T & -2.2 & 34 & 925 -- 4850 & 4 & Peak \\
 47 &  B1952+29  & 0.427 & T & --- & 30 & 610 -- 1418 & 2 & Peak \\
 48 &  B2002+31  & 2.111 & T & -1.5 & 49 & 610 -- 1642 & 3 & Total \\
 49 &  B2045-16  & 1.962 & T & -1.7 & 36 & 325 -- 4850 & 7 & Total \\
 50 &  B2111+46  & 1.015 & T & -2.0 & 9 & 408 -- 4850 & 7 & Total \\
 51 &  B2210+29  & 1.005 & T & -1.4 & 41 & 130 -- 1418 & 4 & Total \\
 52 &  B2224+65  & 0.683 & T$_{1/2}$ & -1.7 & 16 & 325 -- 1642 & 7 & Peak \\
 53 &  B2327-20  & 1.644 & T & -2.1 & 60 & 325 - 610 & 2 & Total \\
\enddata
\end{deluxetable}

It is expected that any variations due to scintillation or lack of flux 
calibration will affect all components of the profile in an identical manner 
and hence any properly formed average profile is useful for the spectral 
difference studies. However, the pulsar need to have different component types 
and these studies are viable in core-cone profiles, T and M, and conal profiles
with clearly distinguished inner and outer cones, $_c$Q and $_c$T. Profile 
classification studies were initiated in the pioneering works of 
\cite{ET_R90,ET_R93} and has been expanded in recent years \citep{MR11,BMM19,
OMR19}. We found around 150 pulsars in the literature that were classified as 
one of the four relevant profile types. The spectra studies can be carried out
only if profile measurements are available at two or more well separated 
frequency bands with sufficient temporal resolution to detect the individual 
components. The emission components also show frequency evolution due to 
various effects like radius to frequency mapping, LOS evolution with frequency,
relative spectral difference between components (the topic of our study), etc.,
such that in many pulsars one or more components either merge together or 
vanish at certain frequencies. This left us with 53 pulsars, where profiles 
with distinct component types are available at more than one frequency.

\begin{figure}
\gridline{\fig{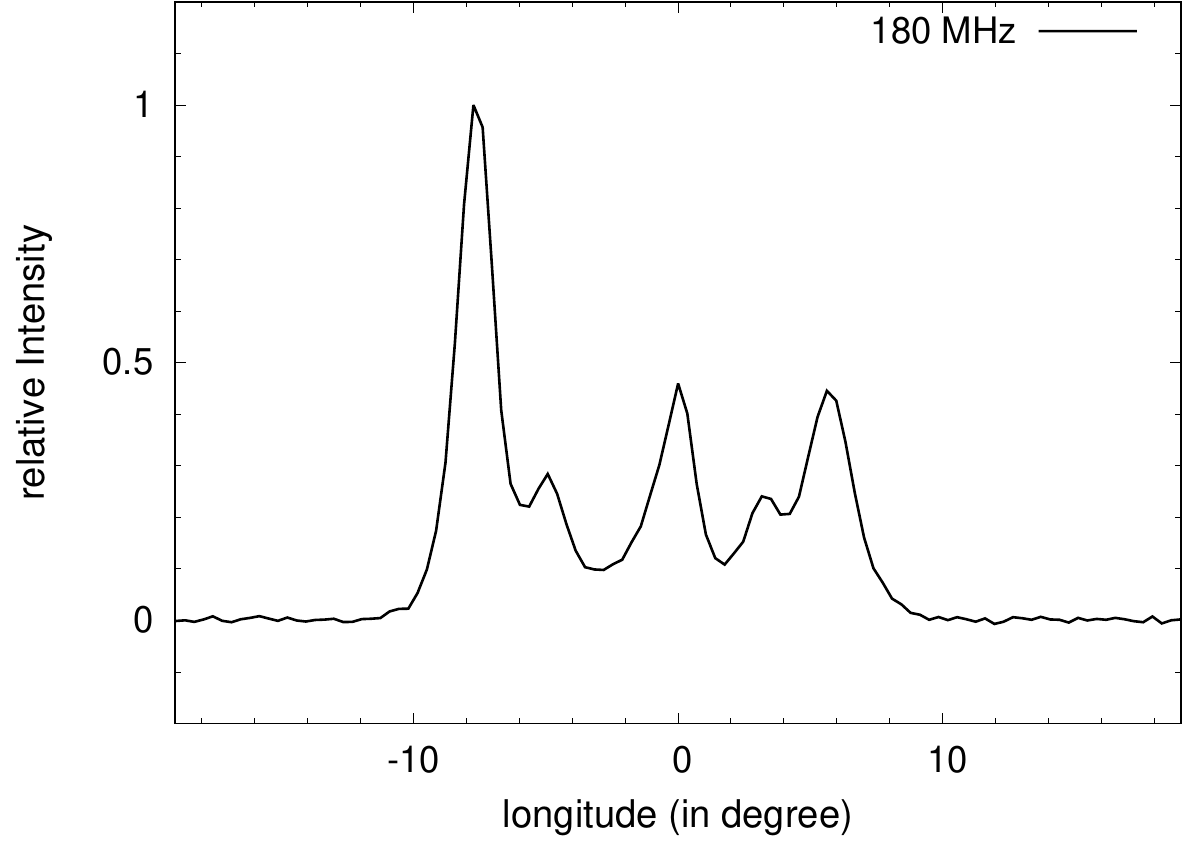}{0.37\textwidth}{}
          \fig{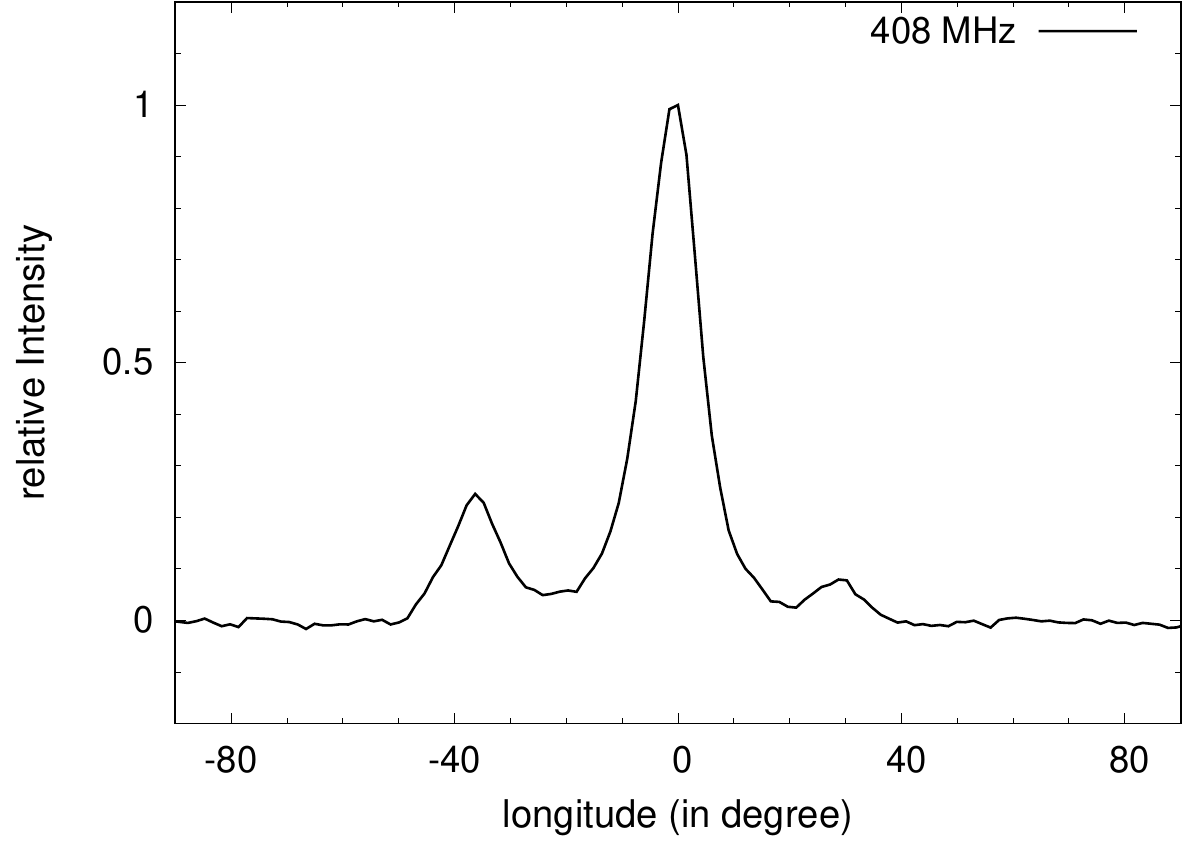}{0.37\textwidth}{}
         }
\gridline{\fig{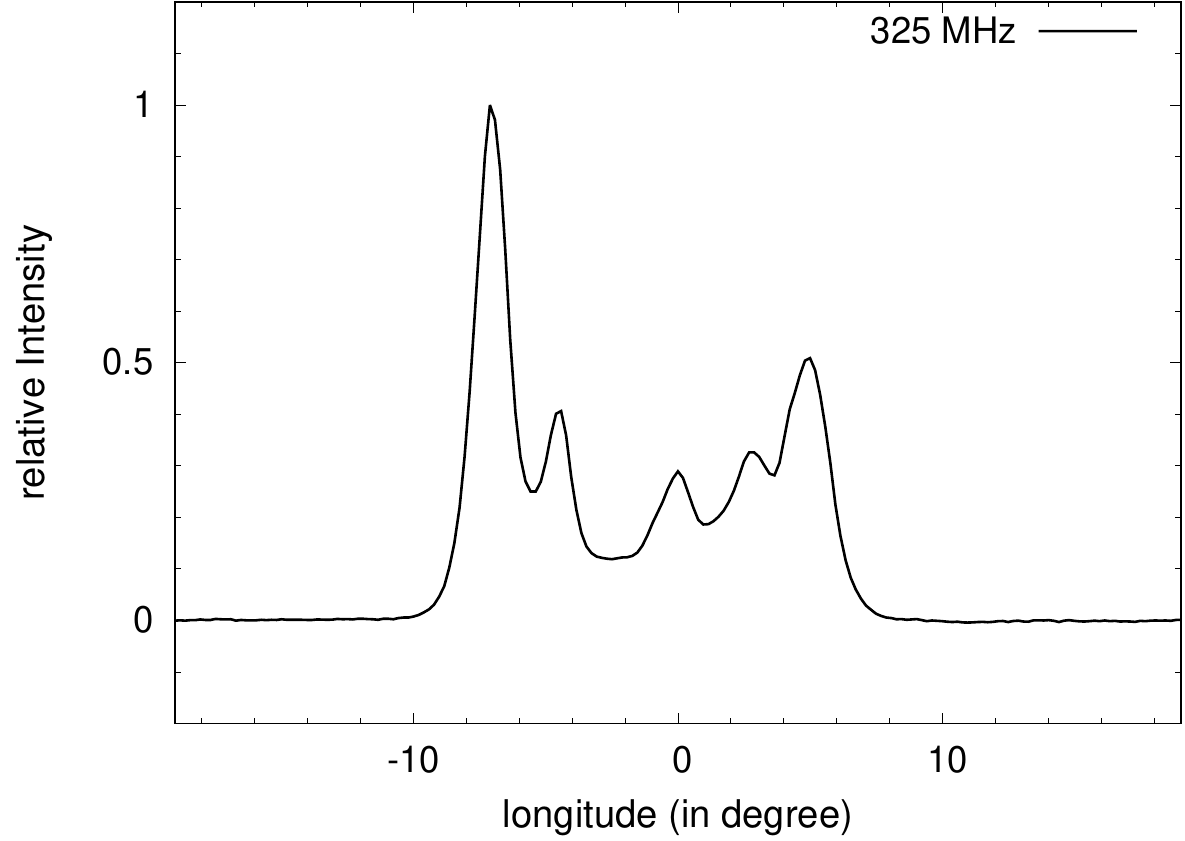}{0.37\textwidth}{}
          \fig{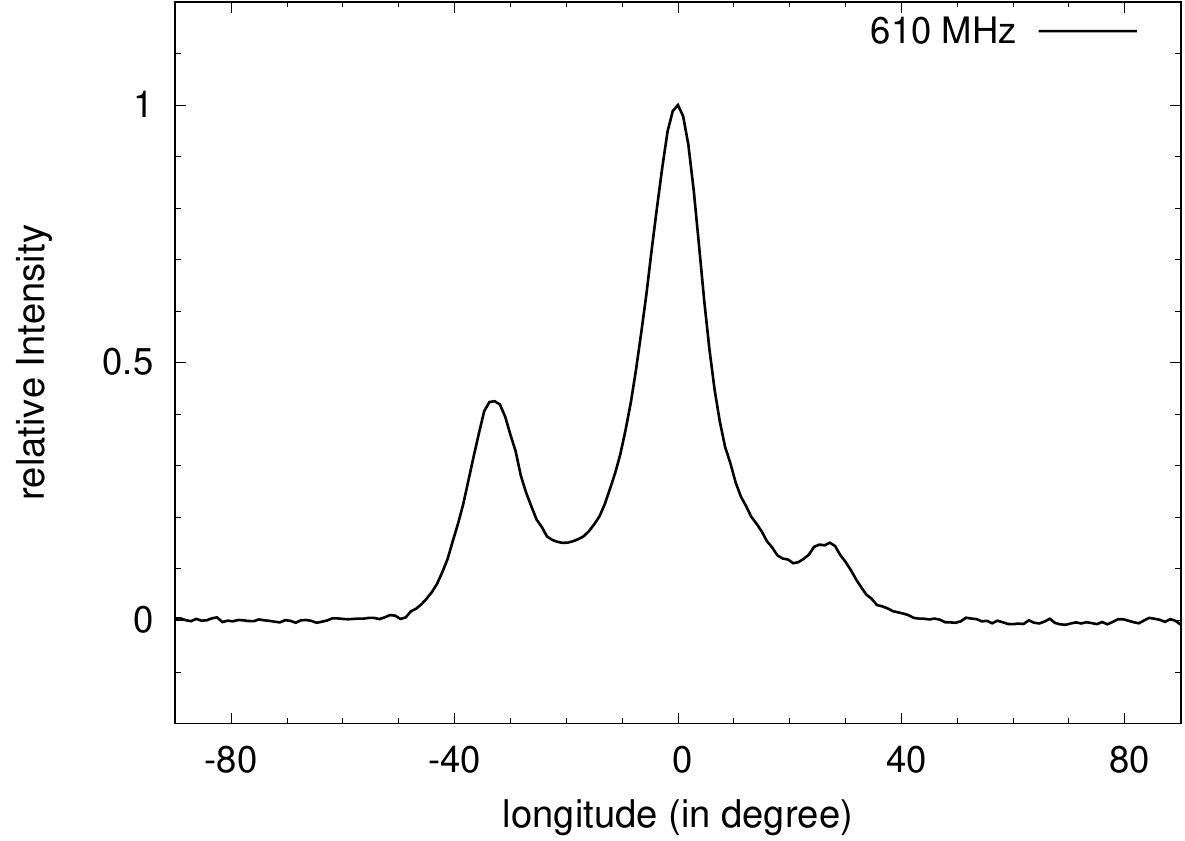}{0.37\textwidth}{}
         }
\gridline{\fig{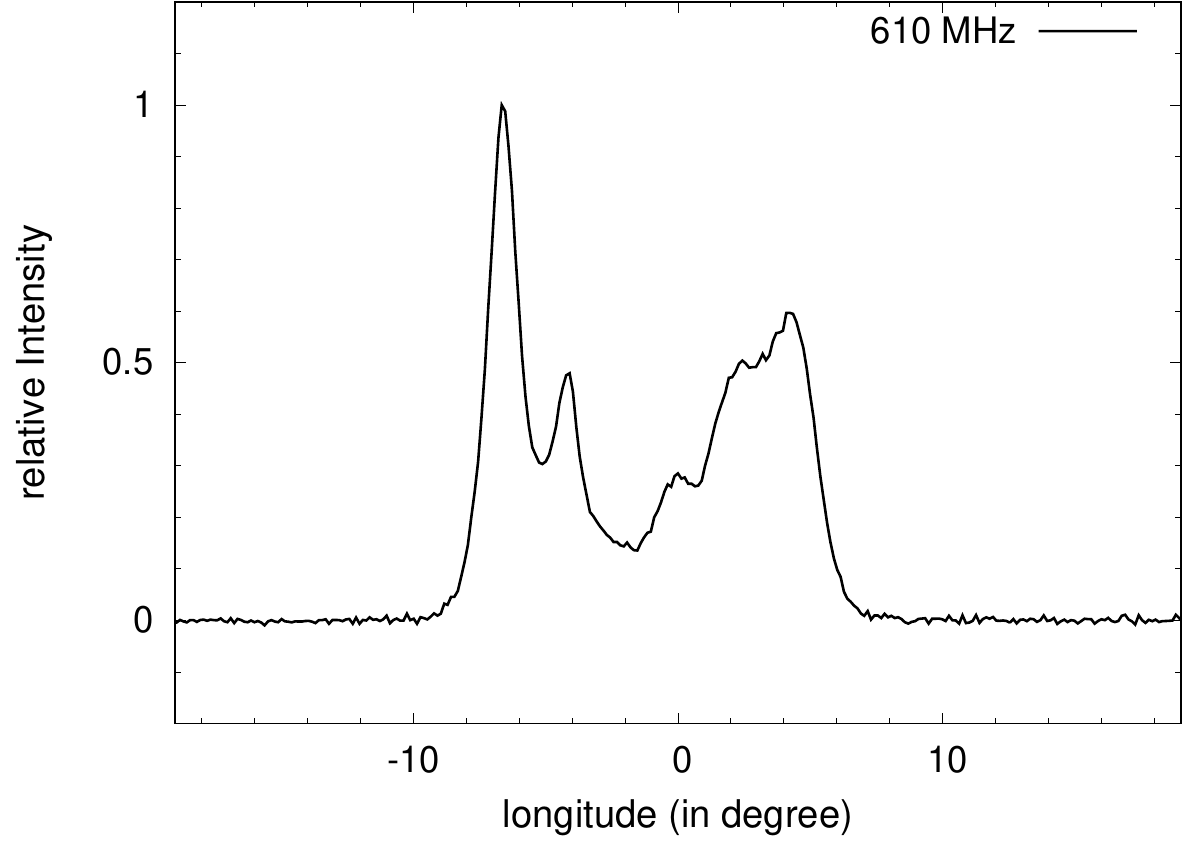}{0.37\textwidth}{}
          \fig{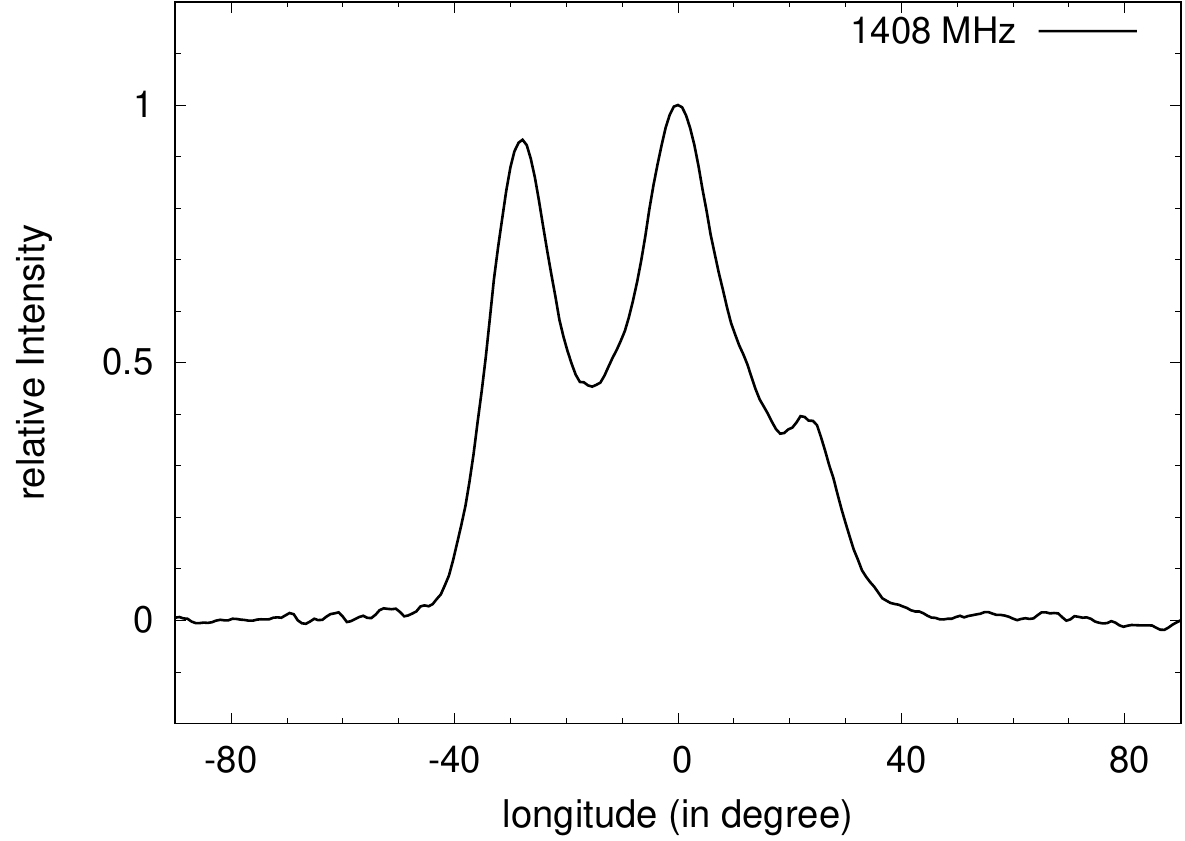}{0.37\textwidth}{}
         }
\gridline{\fig{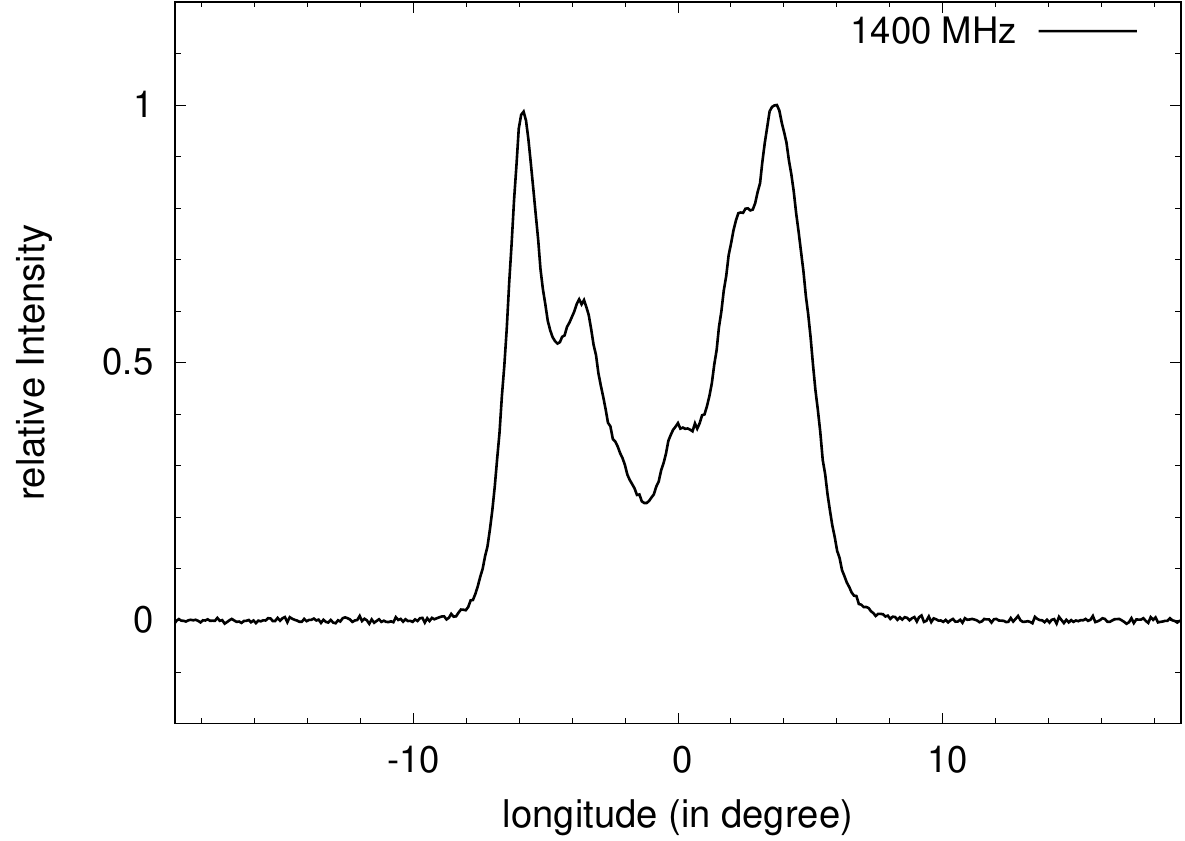}{0.37\textwidth}{}
          \fig{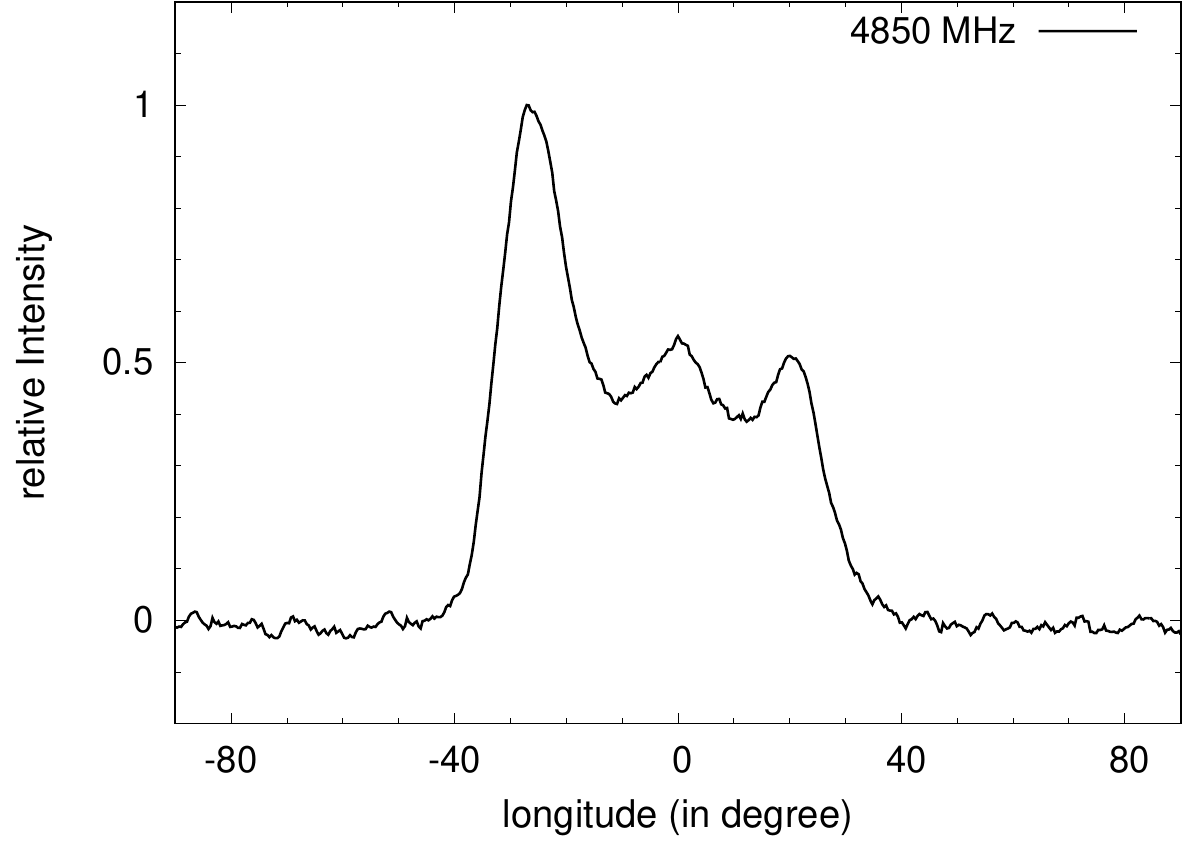}{0.37\textwidth}{}
         }
\caption{The figure shows evolution of the average profile over four different 
frequency bands in two different pulsars, B1237+25 (left panel) with M type 
profile and B2111+46 (right panel) with T profile type. The references for 
each measurement is listed in Table \ref{tab:spect}.
\label{fig:avgprof}}
\end{figure}

Table \ref{tab:srclist} shows details of the 53 pulsars used for the spectra  
studies. The list consists primarily of T and T$_{1/2}$ profiles and include 43
such pulsars, while another 6 are of M type and 4 have $_c$T/$_c$Q profiles. 
The frequency coverage is wide, ranging from around 100 MHz up to 10 GHz in a 
few cases, with around 40 percent (21 pulsars) of the sample having 
measurements at more than three different frequencies. The Table also reports 
the average spectral index for each pulsar, the sample showing typical steep 
spectra behaviour with values ranging mostly between -1.0 and -2.5. The 
presence of the core component in most of the profiles also allowed us to 
estimate the magnetic inclination angle ($\vartheta$), where the 50\% width of 
the core ($W_{50}$) at 1 GHz can be associated with emission geometry as 
$W_{50} = 2.45 P^{-0.5}/\sin{\vartheta}$ \citep{ET_R90}. We also list the 
frequency range over which average profile measurements were available in each 
pulsar as well as the total number of such measurements (Nfreq).

In Fig.\ref{fig:avgprof} two examples of the frequency evolution of the profile
components are shown. The left panel corresponds to the M type pulsar B1237+25
with four profiles at 180 MHz, 325 MHz, 610 MHz and 1400 MHz, while the right 
panel shows the T type pulsar B2111+46 at 408 MHz, 610 MHz, 1408 MHz and 4850 
MHz. In both cases the central core component has higher relative intensities
at the lower frequencies compared to the cones. It is apparent from the figure
that the different component types show significant evolution with frequency
and in many cases the components merge together at one or more frequencies (see
core emission at 1400 MHz for B1237+25). Hence, it becomes difficult to 
estimate the total power in the components for the spectra calculations at 
certain frequencies, and the peak or the average intensities in the components 
provide better estimates in such cases. The last column in Table 
\ref{tab:srclist} lists the method by which the intensities of the components 
were estimated in each pulsar. The peak intensities are estimated from Gaussian
approximation of suitably selected points around the expected peak. In 
instances where the average intensities are used, the adjacent components are 
not clearly separated. The longitude corresponding to the minimum intensity 
between these components is considered the boundary for the measurements. The
intensity ratio between the components is used to estimate the relative 
spectra. As a result no additional scaling is required when using the peak or 
the average intensities since they are expected to affect both components in a 
similar manner.

\section{Relative Spectral difference Measurements}\label{sec:compspec}

\begin{figure}
\epsscale{0.6}
\plotone{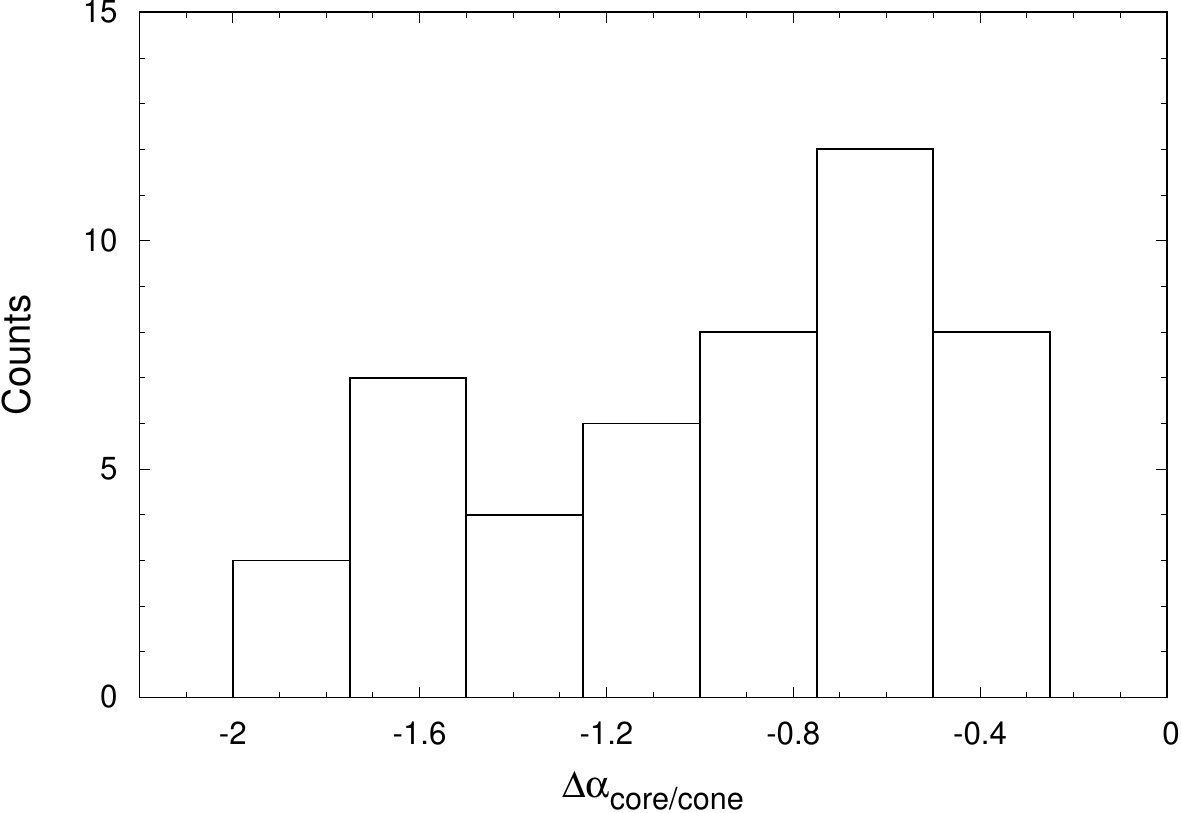}
\caption{The distribution of the spectral difference between the core and conal
components. \label{fig:spectdist}}
\end{figure}

\begin{figure}
\gridline{\fig{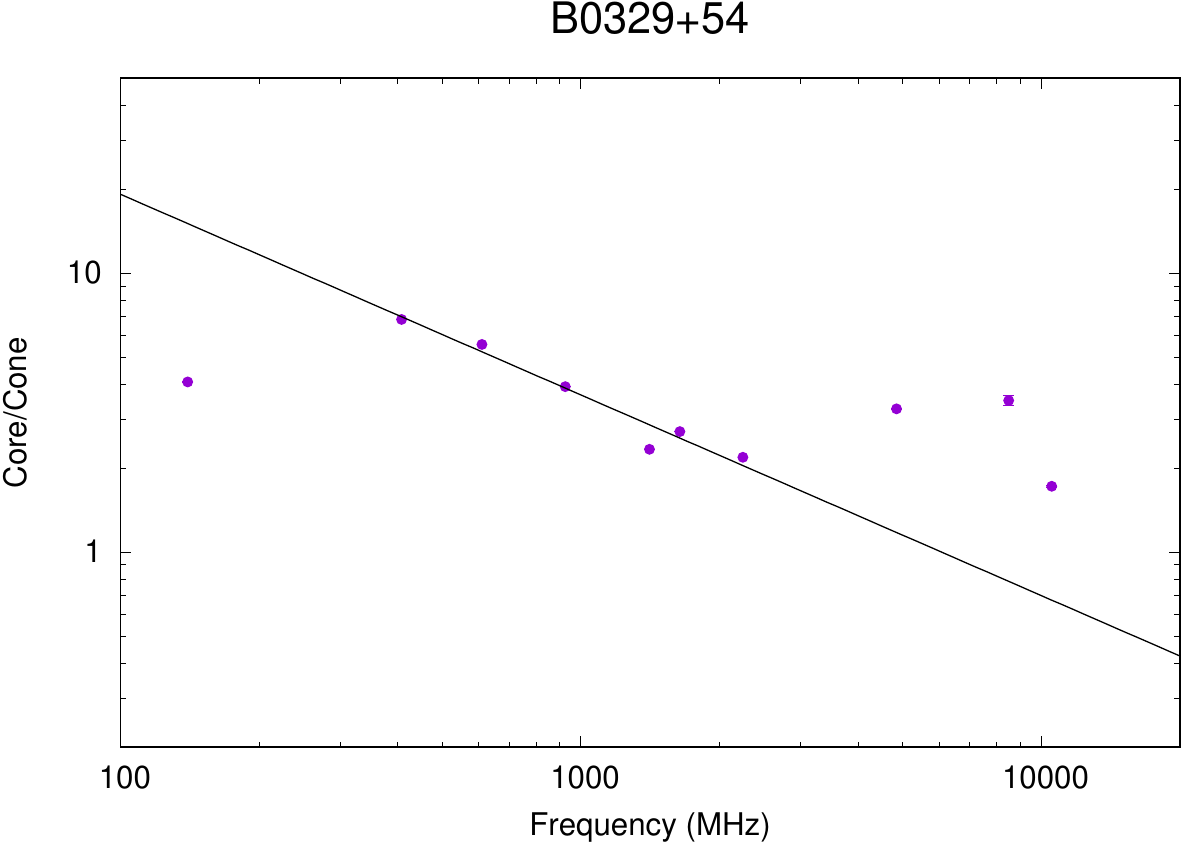}{0.38\textwidth}{}
          \fig{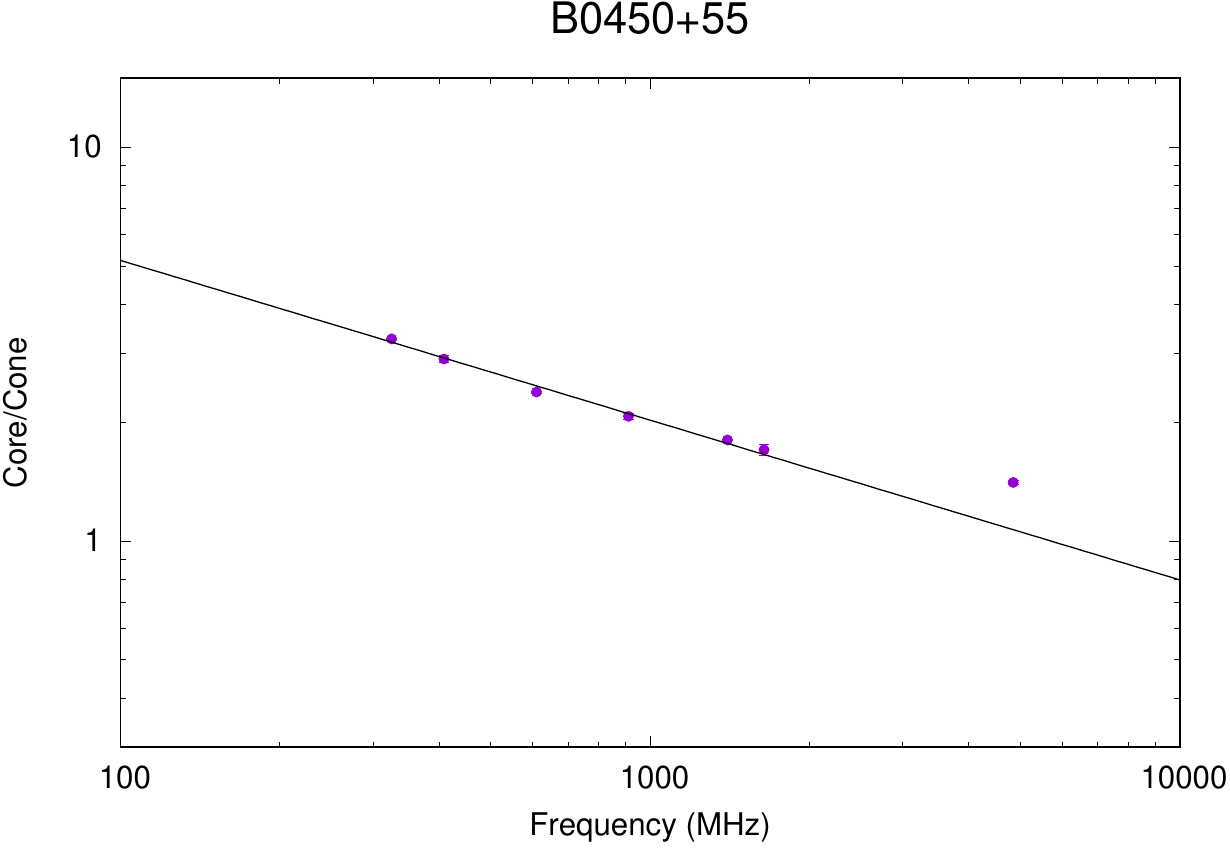}{0.38\textwidth}{}
         }
\gridline{\fig{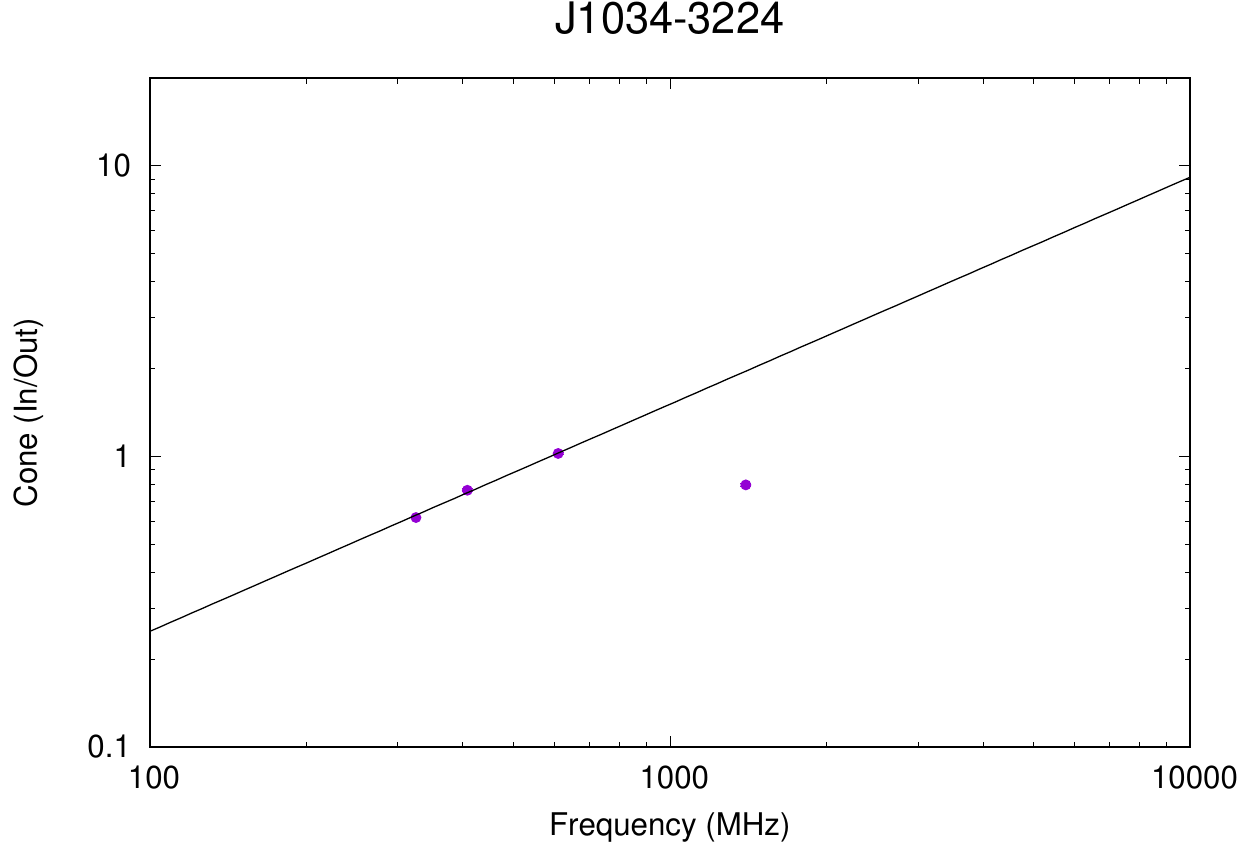}{0.38\textwidth}{}
          \fig{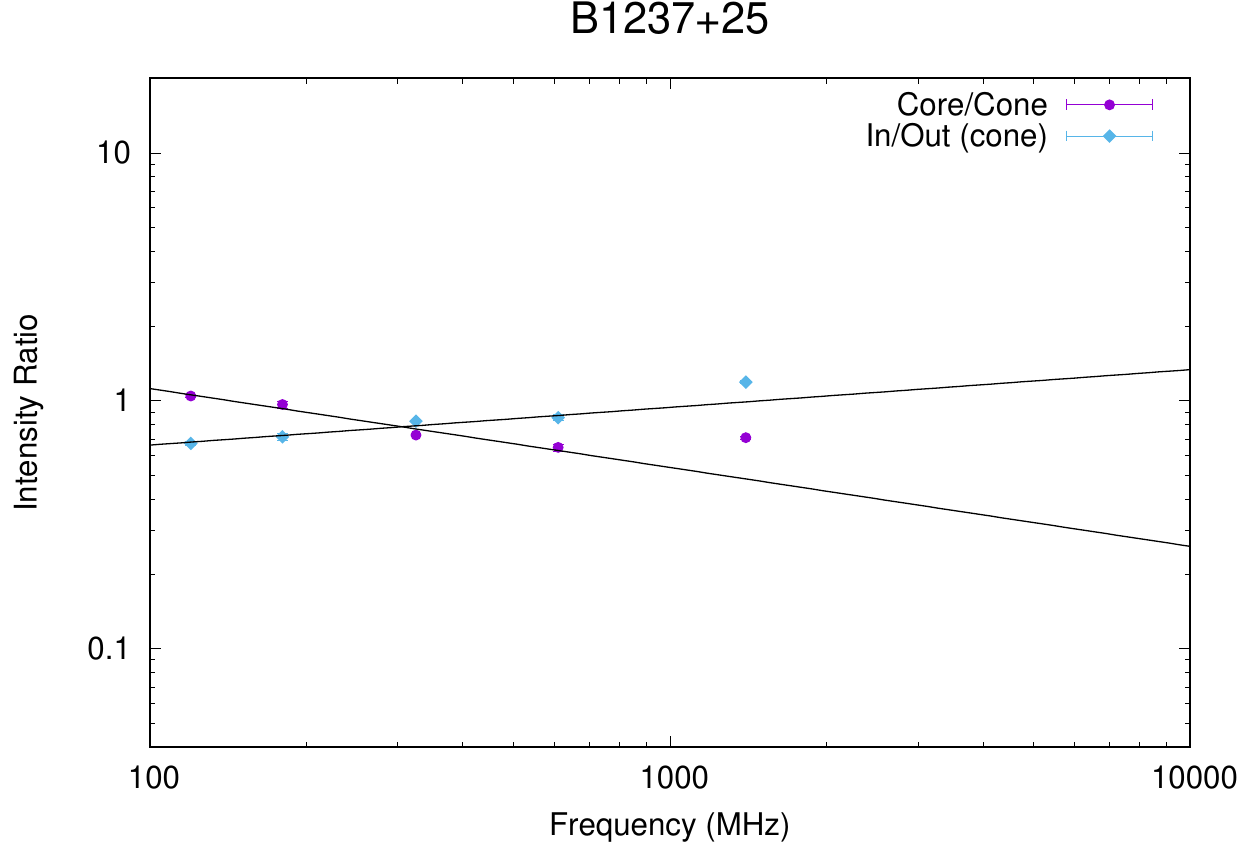}{0.38\textwidth}{}
         }
\gridline{\fig{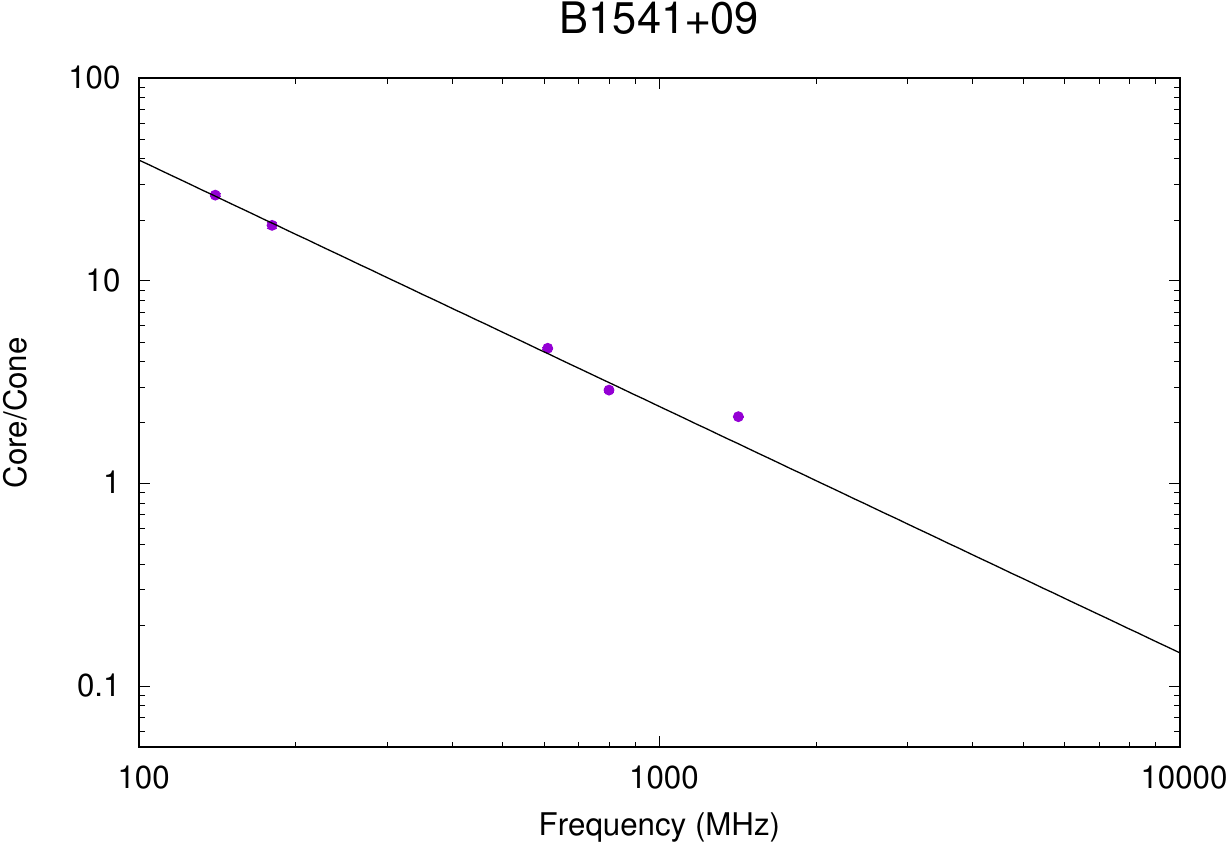}{0.38\textwidth}{}
          \fig{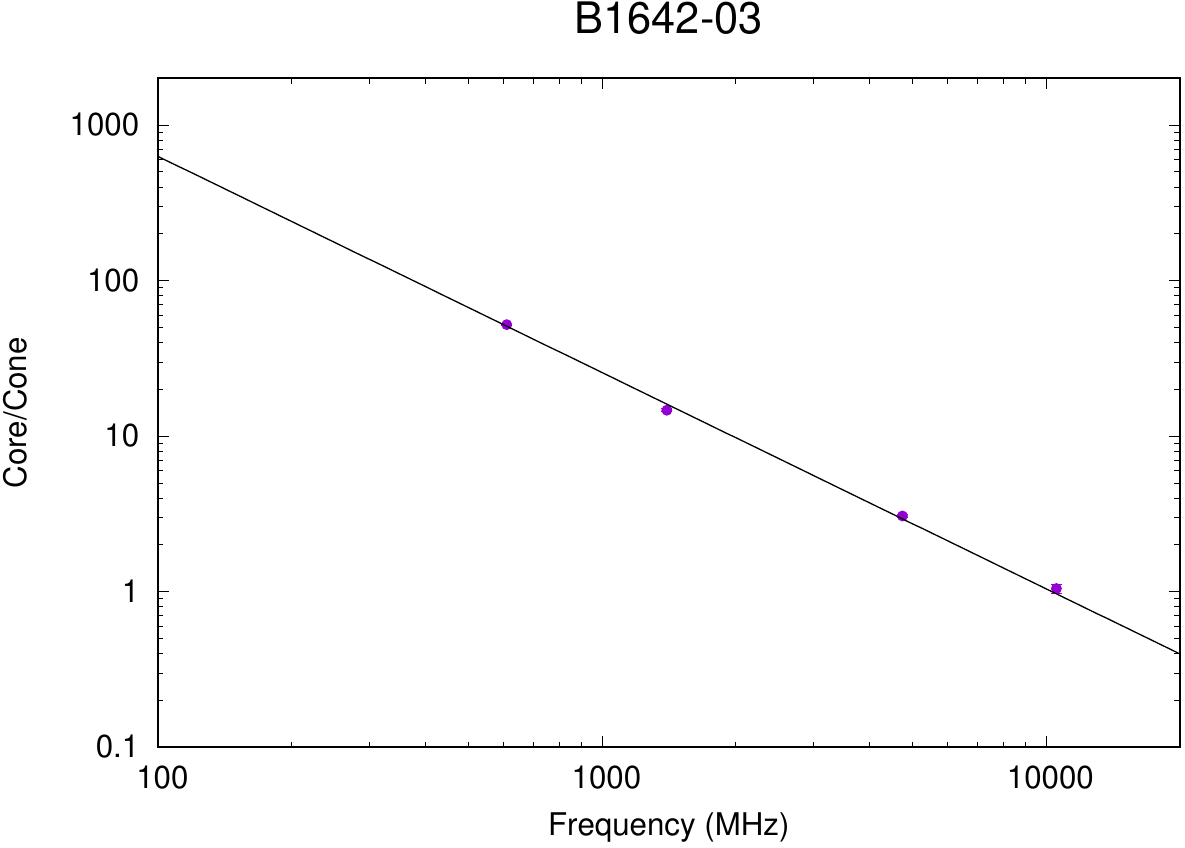}{0.38\textwidth}{}
         }
\gridline{\fig{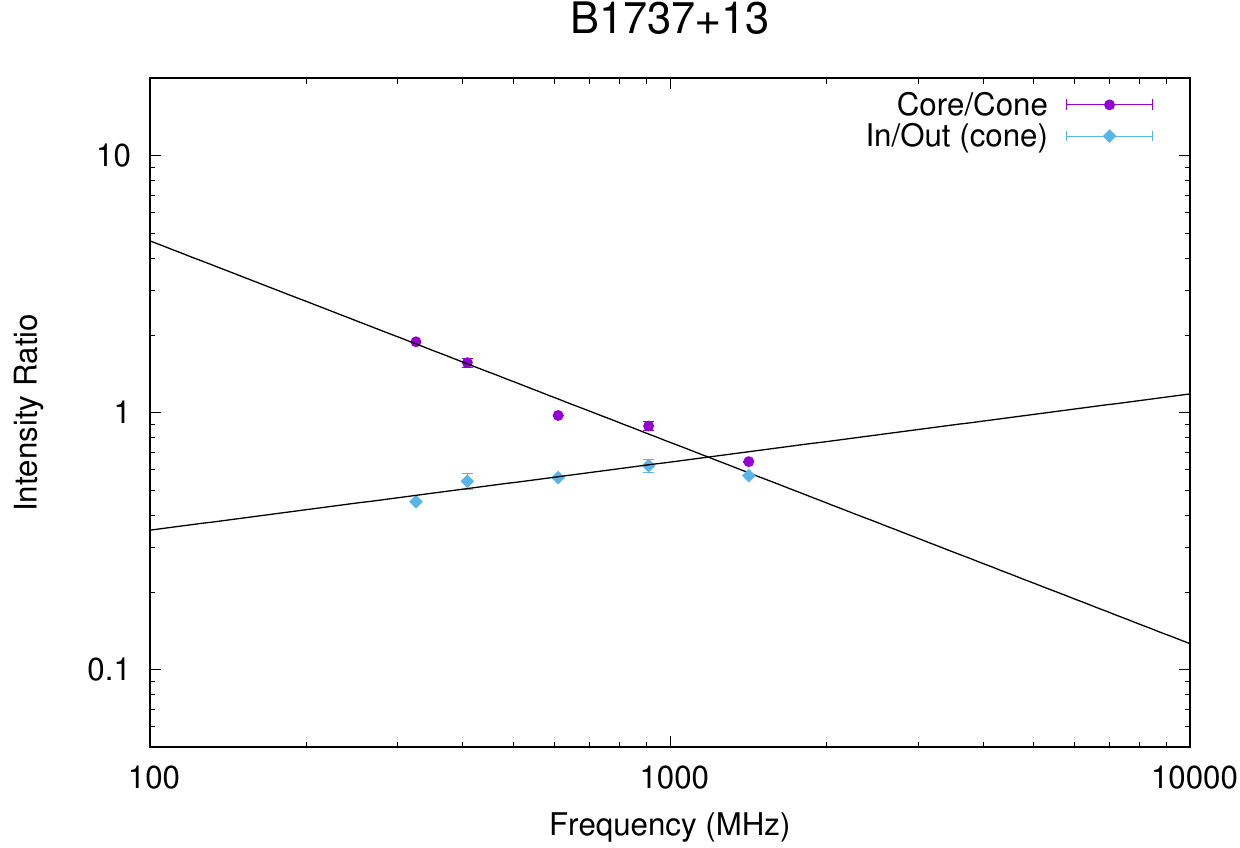}{0.38\textwidth}{}
          \fig{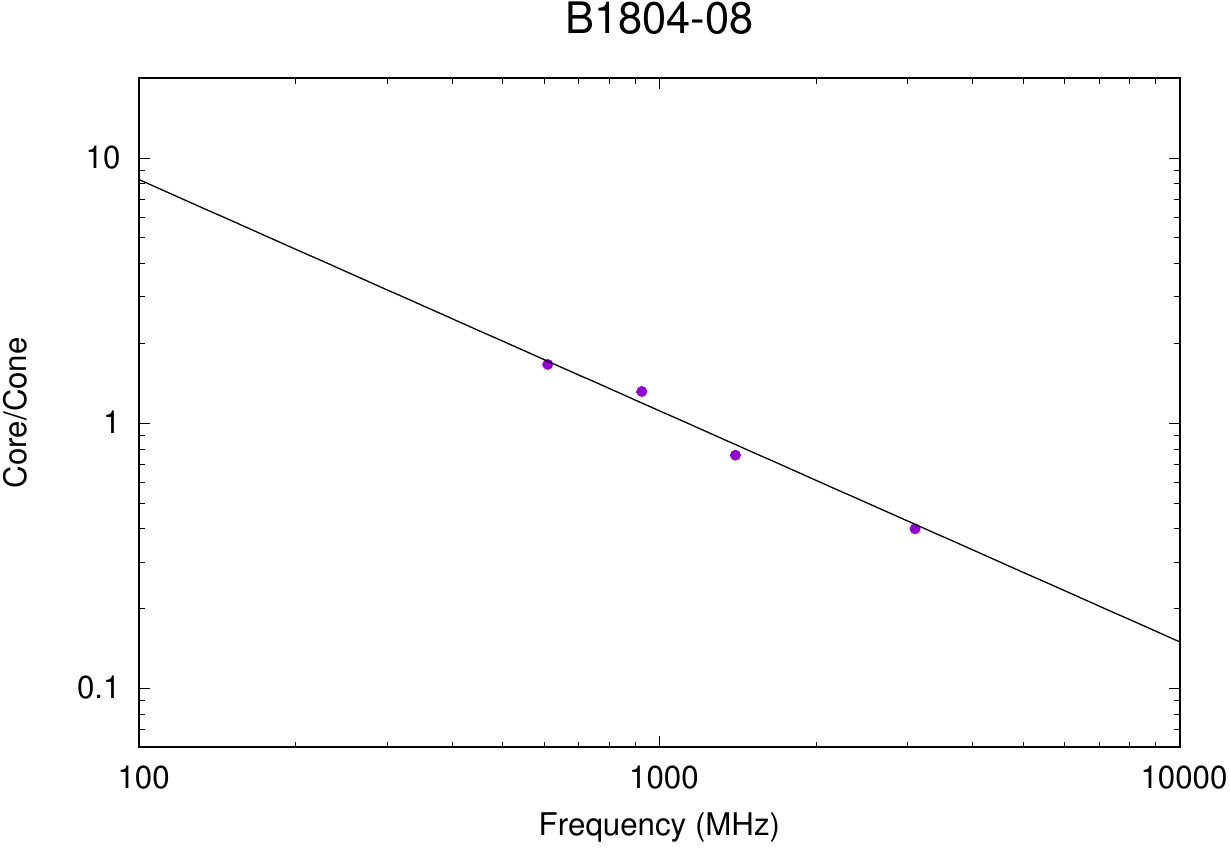}{0.38\textwidth}{}
         }
\caption{The figure shows the spectra of the relative intensities between the 
core and conal components within the pulsar profile in 8 pulsars between 100 
MHz and 10 GHz. The spectral evolution between the inner and outer cones are 
also estimated in PSR B1237+25 and B1737+13 with M profile type as well as PSR 
J1034-3224 with $_c$Q profile. \label{fig:spect1}}
\end{figure}

\begin{figure}
\gridline{\fig{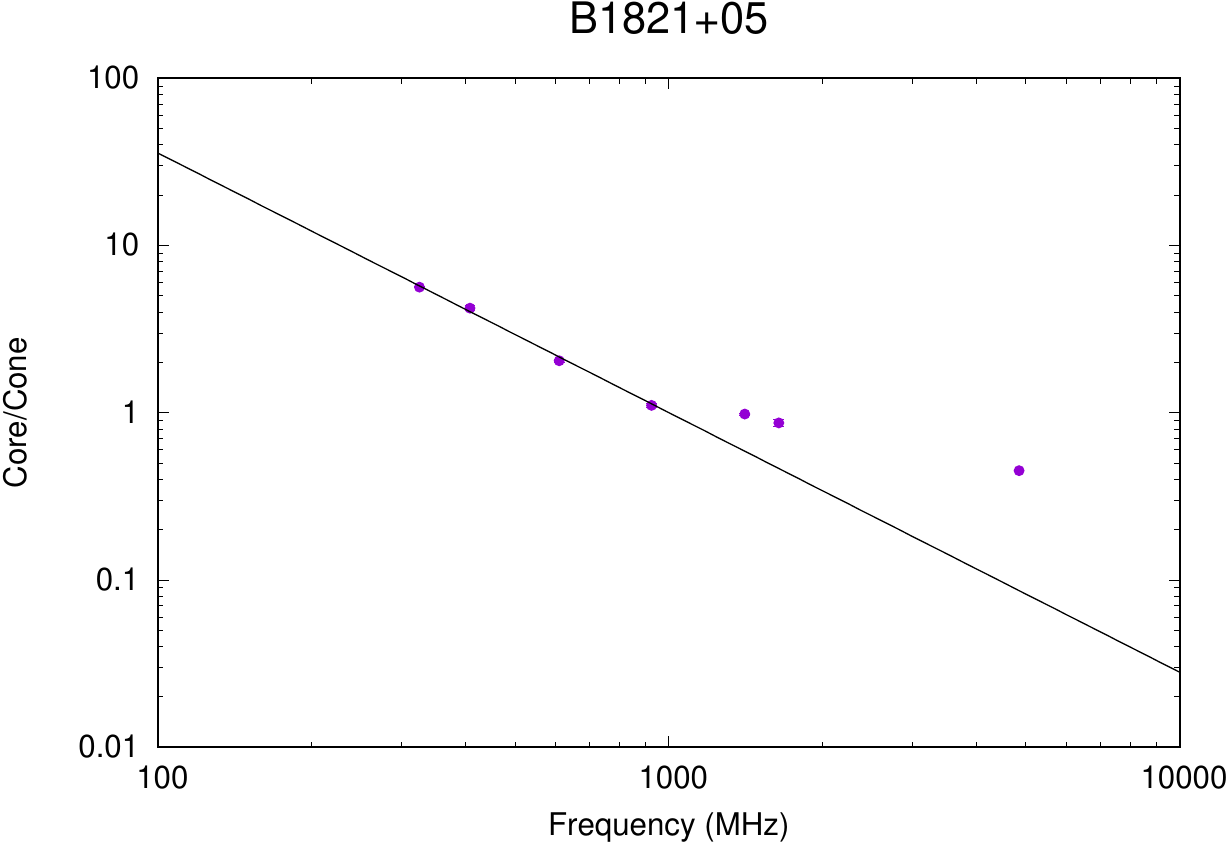}{0.39\textwidth}{}
          \fig{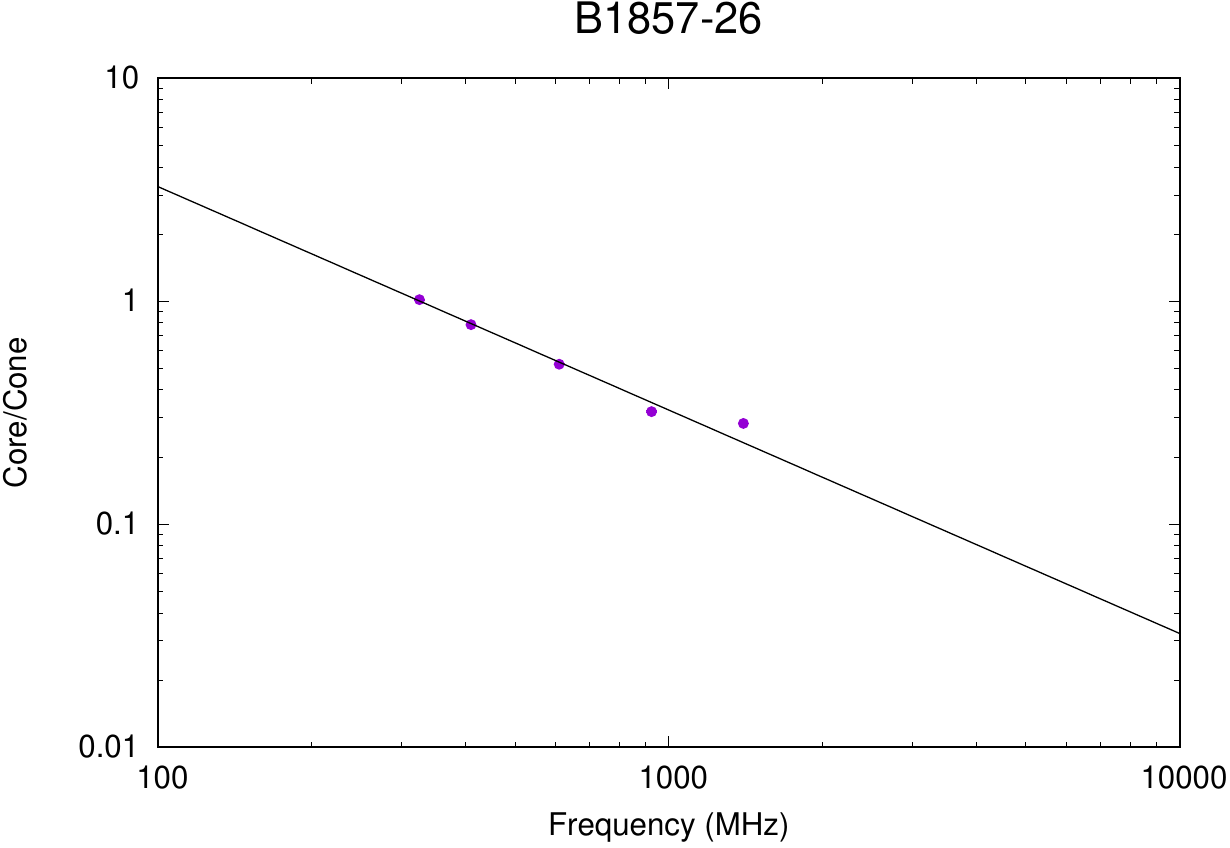}{0.39\textwidth}{}
         }
\gridline{\fig{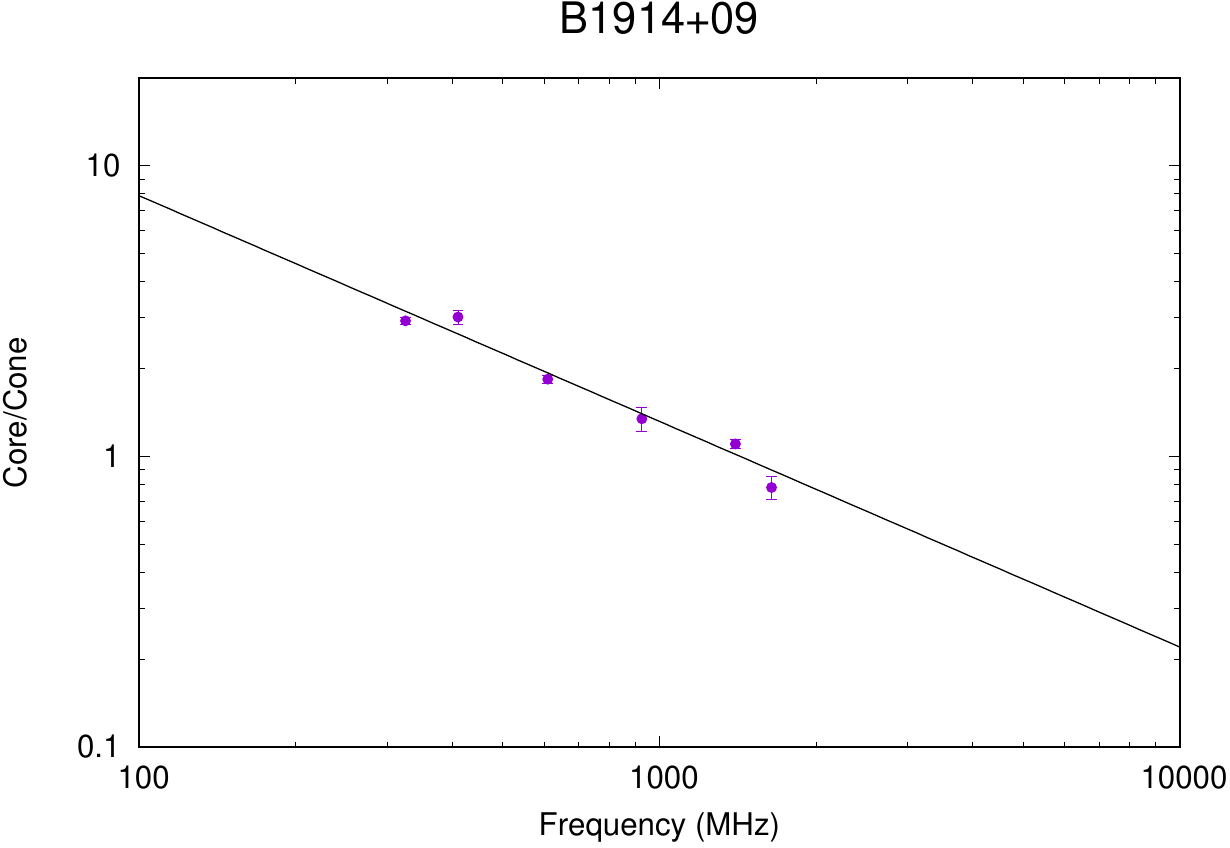}{0.39\textwidth}{}
          \fig{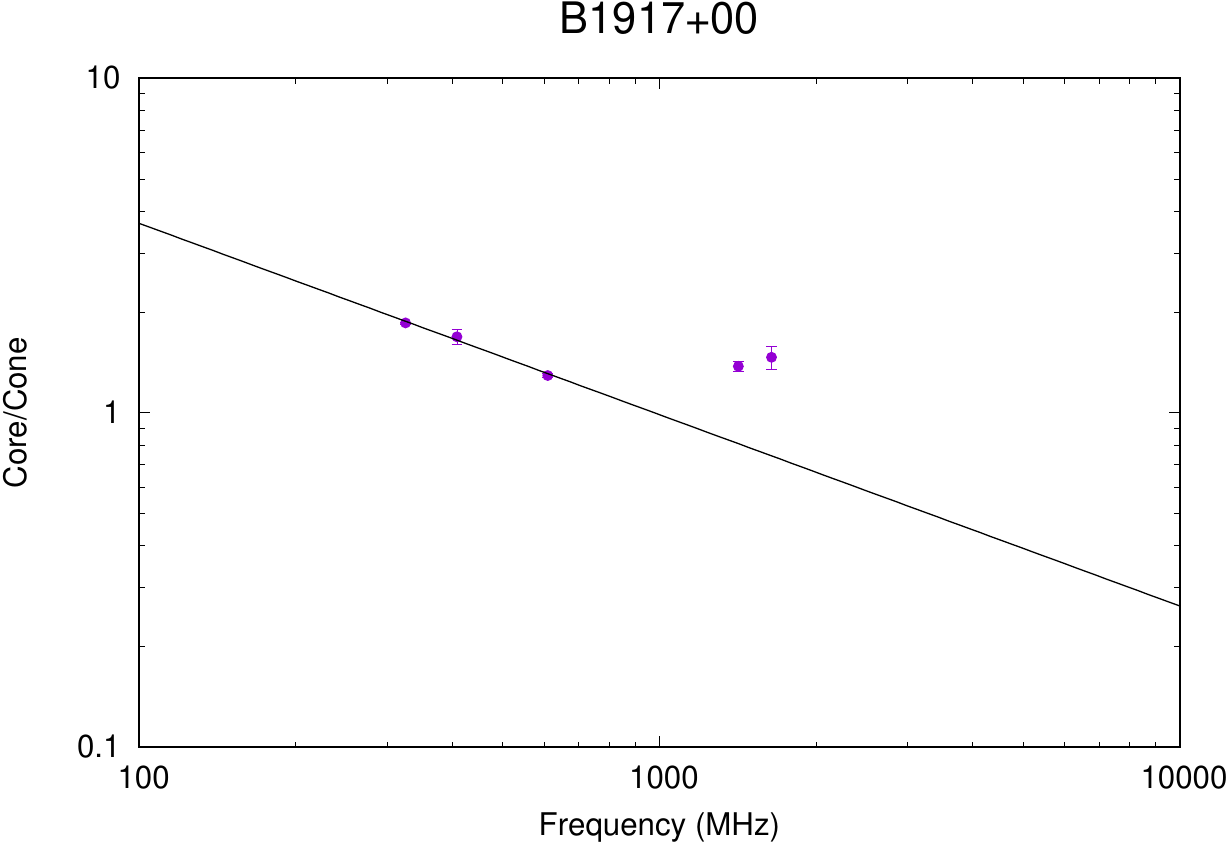}{0.39\textwidth}{}
         }
\gridline{\fig{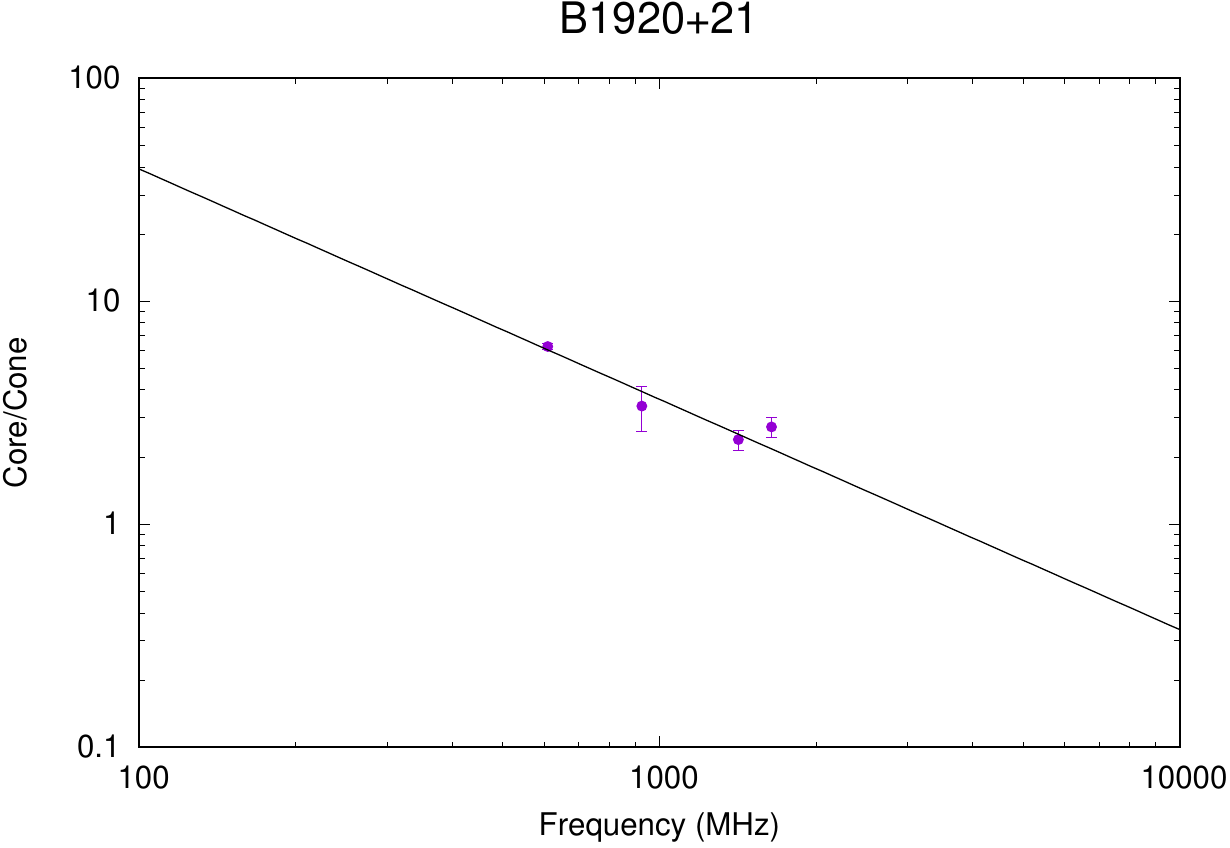}{0.39\textwidth}{}
          \fig{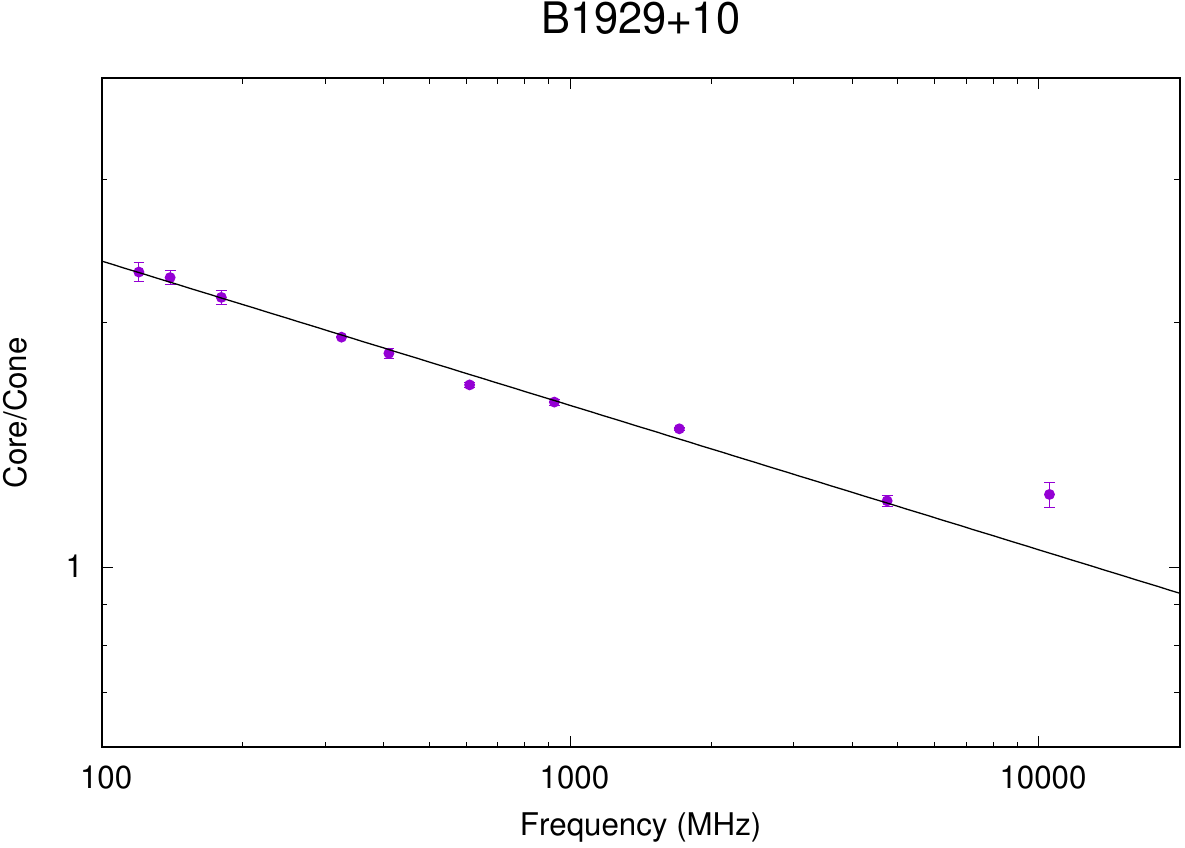}{0.38\textwidth}{}
         }
\gridline{\fig{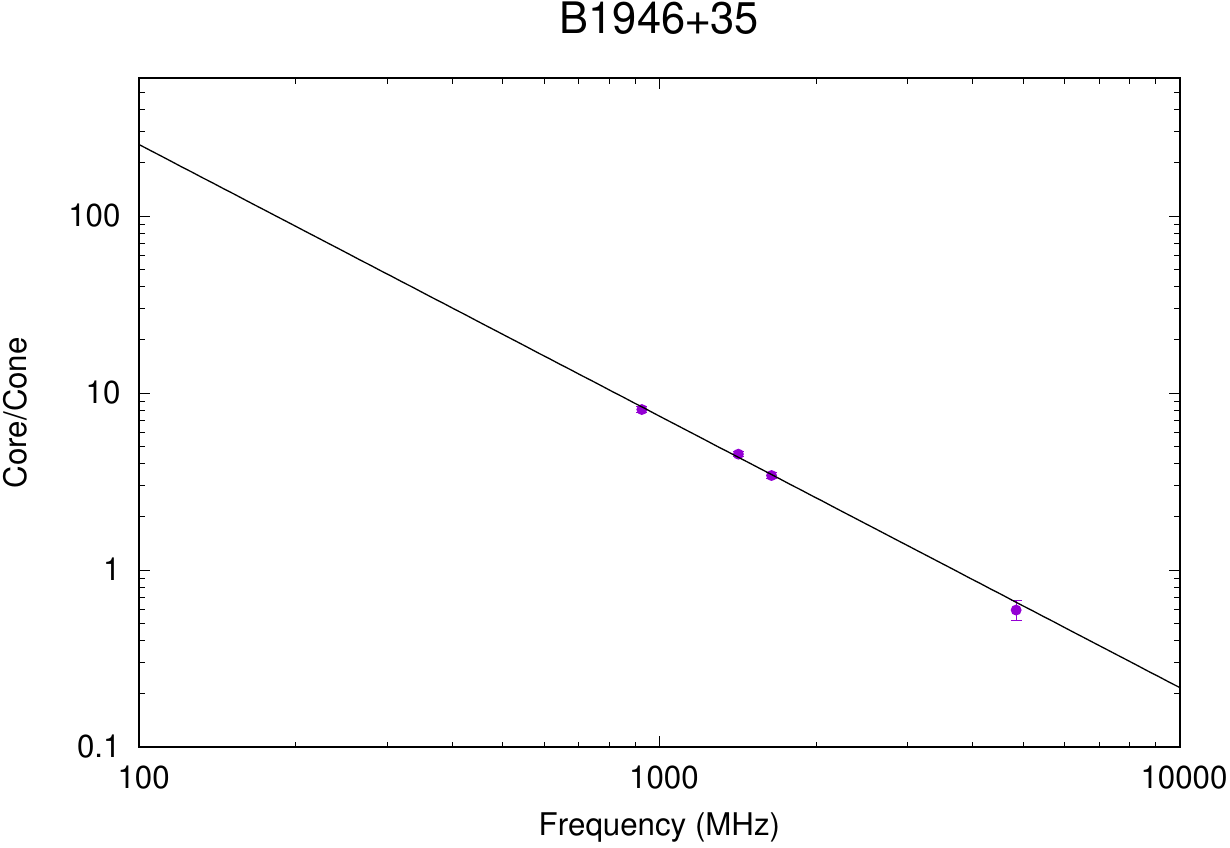}{0.4\textwidth}{}
          \fig{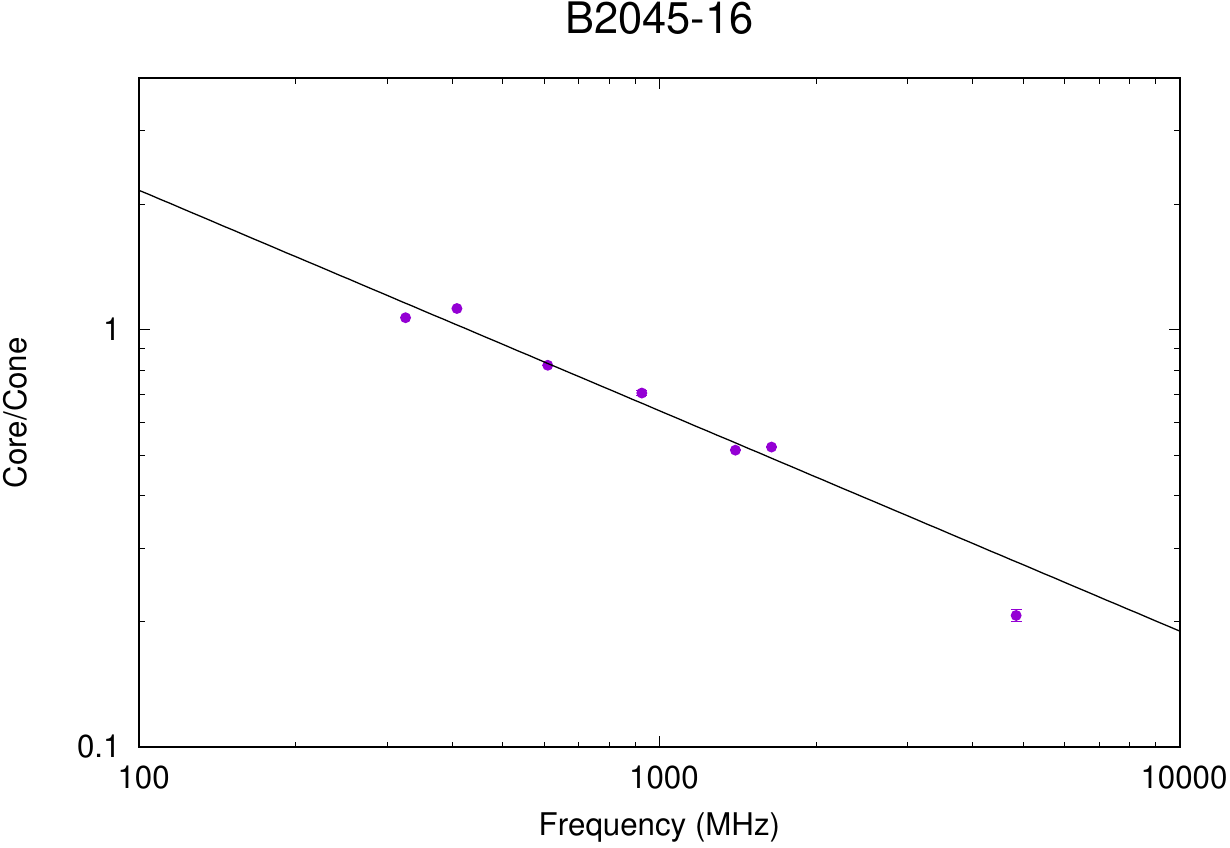}{0.4\textwidth}{}
         }
\caption{The figure shows the spectra of the relative intensities between the 
core and conal components within the pulsar profile in additional 8 other 
pulsars between 100 MHz and 10 GHz. \label{fig:spect2}}
\end{figure}

\begin{figure}
\gridline{\fig{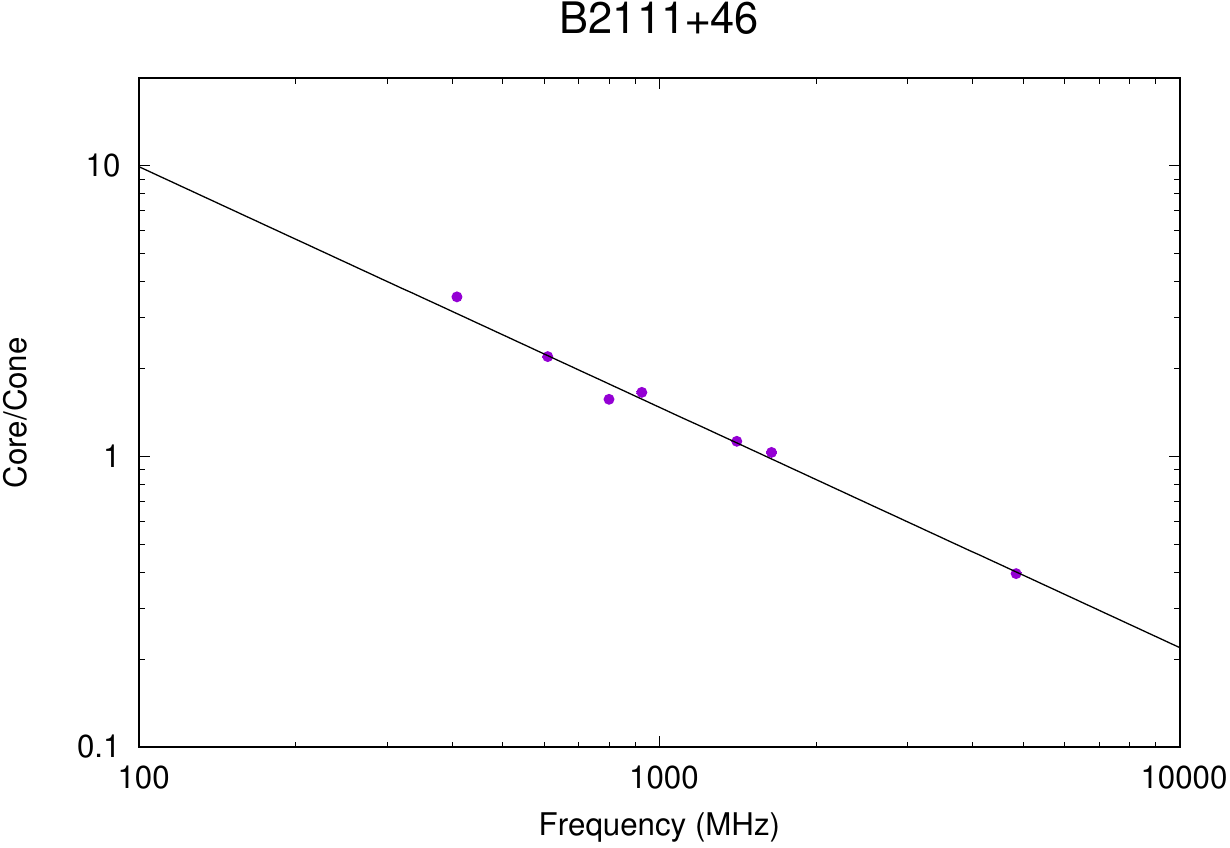}{0.4\textwidth}{}
          \fig{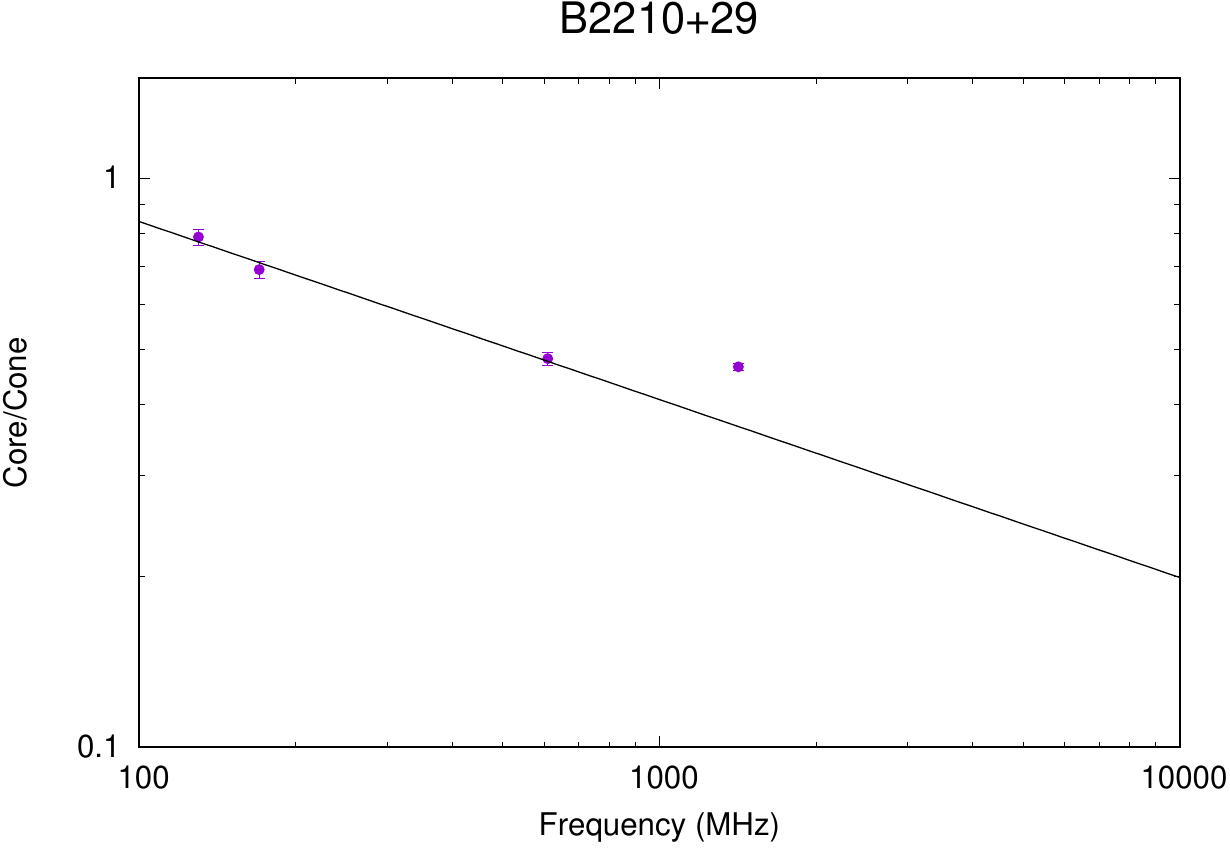}{0.4\textwidth}{}
         }
\gridline{\fig{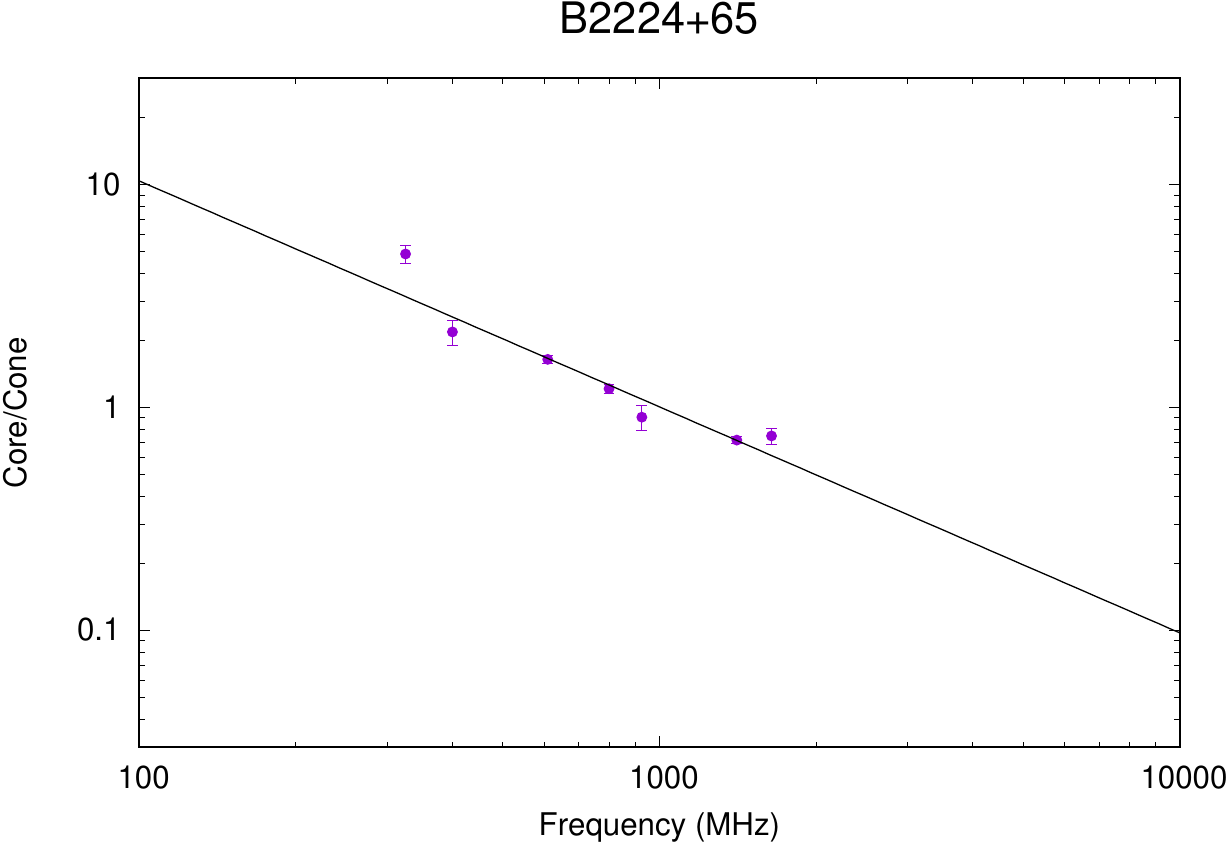}{0.4\textwidth}{}
         }
\caption{The figure shows the spectral nature of the relative intensities 
between the core and conal components within the pulsar profile in 3 pulsars
between 100 MHz and 10 GHz. \label{fig:spect3}}
\end{figure}

We have measured the ratio of the intensities between the core and the cone 
($S_{core}$/$S_{cone}$) in 49 pulsars, the core and inner and outer cones, 
($S_{core}$/$S_{in}$ and $S_{core}$/$S_{out}$) in 5 pulsars, and between the 
inner and outer cones ($S_{in}$/$S_{out}$) in 9 pulsars, from profiles in 
different frequencies. These measurements are reported in Table \ref{tab:spect}
as well as the difference in the spectral index between the different component
types and the frequency range over which the spectra is calculated. The 
spectral difference between the core and conal components 
($\Delta\alpha_{core/cone}$) shows a relatively wide spread between -0.2 and 
-2.0 with a fairly uniform distribution as shown in Fig.\ref{fig:spectdist}. A 
slight peak around $\Delta\alpha_{core/cone}\sim-0.7$ is visible which is 
consistent with earlier results \citep{BMM21}. No clear correlation is seen 
between $\Delta\alpha_{core/cone}$ and different physical parameters like 
period, characteristic age, spin-down energy loss and the average spectral 
index. The spectral difference between the inner and outer cones 
($\Delta\alpha_{in/out}$) in 9 pulsars varies between +0.1 and +0.8 with a mean
value of +0.4. The spectra variation of the relative intensities of different 
components of 19 pulsars where measurements were available over relatively
wide frequency range are shown in Fig. \ref{fig:spect1}, \ref{fig:spect2}, 
\ref{fig:spect3}. We briefly discuss below the profile characteristics and 
component frequency evolution in all these cases. \\

\emph{B0329+54} : The pulsar profile has three distinct components with a 
bright core and has been classified as a T type profile \citep{ET_R93}. The 
pulsar is very bright with average flux density $>$200 mJy at 1 GHz 
\citep{LYL95} and has been observed over a wide frequency range from 40 MHz to 
around 30 GHz. The core component is asymmetrical with elongation near the 
leading edge. The pulsar emission is scattered and the components 
indistinguishable at frequencies below 100 MHz while one or more components 
vanish at the high frequency range above 10 GHz. The components also exhibit 
radius to frequency mapping with the two conal components being well separated 
from the core at lower frequencies, while the components merge at frequencies 
above 1 GHz but are still distinguishable up to 10 GHz. $S_{core}$/$S_{cone}$ 
shows a power law frequency dependence between 400 MHz and 2 GHz (see Fig. 
\ref{fig:spect1}, first row, left panel) with $\Delta\alpha_{core/cone}\sim$
-0.7. The spectral difference are flatter at lower frequencies while at the 
high frequency end the relative intensities become more irregular. \\ 

\emph{B0450+55} : The pulsar has an asymmetric profile with three merged 
components with trailing component showing a prominent tail. The profile was
classified as T type and is part of the partial cone sample \citep{MR11}. The 
leading component component is not distinguishable from the core at frequencies
below 300 MHz and as result these profiles not used. Due to the components 
being merged together the average intensities are used to estimated the 
evolution of the relative spectra between the components which show a power law
dependence between 300 MHz and 4.8 GHz with $\Delta\alpha_{core/cone}\sim$ -0.4
(see Fig. \ref{fig:spect1}, first row, right panel). \\

\emph{J1034-3224} : The pulsed emission is relatively wide occupying around 30 
percent of the profile window and can be classified into two distinct parts the
main pulse and a precursor component \citep{BM18a}. The precursor becomes more 
prominent at higher frequencies ($>$ 1 GHz) and is characterised by large 
fractional linear polarization and flat PPA traverse across the component 
\citep{BMR15}. The main pulse consists of four distinct conal components and is 
classified as a $_c$Q profile type. The pulsar was observed between 300 MHz and
1.4 GHz and the main pulse shows distinct evolution over this range. A large 
plateau is seen between the inner conal components at 325 MHz. The separation 
between the inner cones narrows with frequency and at 1.4 GHz the two 
components merge. The intensity of the outer cones increases in comparison with
the inner cones with decreasing frequency and has a relative difference in 
spectral index below 1 GHz as $\Delta\alpha_{in/out}\sim$ +0.78 (see Fig. 
\ref{fig:spect1}, second row, left panel). The 1.4 GHz measurements is not used for the relative spectral index estimate. \\

\emph{B1237+25} : The central core in this M type profile is relatively weak 
and is clearly visible only at frequencies of 325 MHz and below (see Fig. 
\ref{fig:avgprof}). The spectra of the relative intensities between the core 
and the cones show a power law dependence below 1 GHz (see Fig. 
\ref{fig:spect1}, second row, right panel), while at higher frequencies the 
measurements become less clear due to the core merging with the inner cone and 
the conal components overlapping each other. As a result only the leading side
of the core is used to estimate the relative intensities at all frequencies. In
this pulsar the spectral differences are relatively flat with
$\Delta\alpha_{core/cone}\sim$ -0.3 while $\Delta\alpha_{in/out}\sim$ +0.15, 
which are near the upper and lower ends of the distributions, respectively. \\

\emph{B1541+09} : The pulsar has a wide profile which encompasses almost half
the rotation period and has a T type classification. The conal emission is very
low at lower frequencies from 400 MHz and below, but becomes more prominent at 
higher frequencies. The relative intensities show a power law spectra from 100 
MHz to 1.4 GHz with relatively steep spectral difference 
$\Delta\alpha_{core/cone}\sim$ -1.2 (see fig. \ref{fig:spect1}, third row, 
left panel). \\

\emph{B1642-03} : The pulsar is classified as T type profile with a prominent 
core emission and has measurements over a wide frequency range between 100 
MHz and 10 GHz. However, the conal emission is relatively weak and undetectable
at frequencies below 600 MHz. At higher frequencies the conal emission merges 
with the core without clear boundaries between them and hence the average 
intensities are used for estimating the relative spectra evolution. The 
estimated spectra between 610 MHz and 10 GHz is $\Delta\alpha_{core/cone}\sim$ 
-1.4 (see Fig. \ref{fig:spect1}, third row, right panel). \\

\emph{B1737+13} : The pulsar profile has a prominent core emission with the 
profile classified as M type. However, the pulsar profile is asymmetric with 
only one conal component seen in the trailing side for most of the observing 
frequencies. The core emission shows a steeper spectra compared to the cone 
with the relative intensities showing a power law nature between 325 MHz and 
1.4 GHz with $\Delta\alpha_{core/cone}\sim$ -0.8 (see Fig. \ref{fig:spect1}, 
fourth row, left panel). The outer cones also exhibit a steeper spectra 
compared to the inner cone with $\Delta\alpha_{in/out}\sim$ +0.3. \\

\emph{B1804-08} : The pulsar has a T type profile with multiple reported 
measurements between 400 MHz and 5 GHz. At lower frequencies less than 900 MHz
the components are merged together and cannot be measured separately. Between 
900 MHz and 3 GHz the relative intensities show a power law spectra (see Fig. 
\ref{fig:spect1}, fourth row, right panel) with $\Delta\alpha_{core/cone}\sim$
-0.9. The components are once again indistinguishable in 5 GHz profile. \\

\emph{B1821+05} : The pulsar also has a T profile type with measurements 
available from 100 MHz to 5 GHz. The conal emission is not seen around 100 MHz 
while at frequencies above 1 GHz the core merges with the trailing cone. The 
relative intensities between the core and conal components show a power law 
spectra between 300 MHz and 1 GHz with spectral index 
$\Delta\alpha_{core/cone}\sim$ -1.55 (see Fig. \ref{fig:spect2}, first row, 
left column). The spectra becomes flatter at the higher frequency range likely 
due to mixing of component intensities. As a result the measurements at 
1.4 GHz, 1.6 GHz and 4.8 GHz are not used for the relative spectral index 
estimate. \\

\emph{B1857-26} : This pulsar with M type profile shows prominent core as well 
as pairs of inner and outer cones between 300 MHz and 1 GHz. The components 
merge together at higher frequencies, and although it is still possible to 
distinguish the core at 1.4 GHz, the inner and outer cones merge together and 
cannot be measured separately. The evolution of the spectra of relative 
intensity between the core and the cones can be constrained using a power law 
nature between 300 MHz and 1.4 GHz with $\Delta\alpha_{core/cone}\sim$ -1.0 
(see Fig. \ref{fig:spect2}, first row, right column). The outer cone has a 
steeper spectra than the inner cone between 300 MHz and 1 GHz with 
$\Delta\alpha_{in/out}\sim$ +0.35. \\

\emph{B1914+09} : The pulsar has two components in the profile with the leading
component being the core and the trailing component the cone, and classified as
T$_{1/2}$. The components merge together and cannot be distinguished at 
frequencies around 100 MHz and 5 GHz. In the intervening frequency range of 325
MHz and 1.4 GHz the relative intensity of the core and the conal components 
shows a power law spectral dependence with $\Delta\alpha_{core/cone}\sim$ -0.8 
(see Fig. \ref{fig:spect2}, second row, left column). \\

\emph{B1917+00} : The pulsar is characterised by a T type profile with the 
conal components not clearly separated from the core. In many profiles the time
resolutions are insufficient to resolve the components. The relative 
intensity between the core and the cone show a power law dependence between 325
MHz and 610 MHz which flattens at 1.4 GHz (see Fig. \ref{fig:spect2}, second 
row, right column). The estimated $\Delta\alpha_{core/cone}$ between 325 and 
610 MHz is around -0.6. The high frequency measurements at 1.4 GHz and 1.6 GHz 
are not used for the relative spectral index estimate. \\

\emph{B1920+21} : The pulsar has a T$_{1/2}$ profile with a prominent core 
emission along with a trailing conal component. At frequencies around 100 MHz 
the profile is affected by scattering. Below 600 MHz the conal emission is very
weak and cannot be clearly measured, while between 600 MHz and 1.6 GHz the 
relative intensities between the core and the cone has a power law spectral 
nature with $\Delta\alpha_{core/cone}\sim$ -1.0 (Fig. \ref{fig:spect2}, third 
row, left column). \\

\emph{B1929+10} : The pulsar has a diverse profile where in addition to the 
main pulse an interpulse emission is seen approximately 180$\degr$ away in 
phase. There is also the presence of a wide postcursor component between the 
main pulse and the interpulse \citep{RR97,BMR15,KYP21}. The main pulse also 
shows high levels of linear polarization with almost 100 percent fractional 
polarization and hence is often used to calibrate the polarization response in
Telescopes \citep{MBM16}. As a result the pulsar has been extensively observed
over a wide frequency range between 100 MHz and 20 GHz. The main pulse is 
classified as a T type profile with three merged components seen around 1.4 
GHz. At lower frequencies the leading conal component is not clearly visible 
and hence we have used the peak intensities of the central core and the 
trailing cone to estimate the relative spectral evolution (see Fig. 
\ref{fig:spect2}, third row, right column). The relative intensity shows a 
power law spectral dependence between 100 MHz and 8.5 GHz with a comparatively
low $\Delta\alpha_{core/cone}\sim$ -0.2. At 10 GHz a flattening in the spectra
is seen which may be likely due to the merging of the two components as they 
become indistinguishable at still higher frequencies. \\

\emph{B1946+35} : The pulsar emission is affected by scattering and the 
profiles show prominent elongated tails below 900 MHz where the individual 
components cannot be identified. At higher frequencies the presence of a T type
profile is seen with a prominent core emission. The leading conal component is 
merged with the core around 1 GHz but becomes clearly visible at 5 GHz. The 
peak intensities of the core and trailing core have been used to estimate their 
relative spectral evolution. Between 900 MHz and 5 GHz the relative intensity 
show a power law dependence with $\Delta\alpha_{core/cone}\sim$ -1.5 (Fig.
\ref{fig:spect2}, fourth row, left column). \\

\emph{B2045-16} : The profile is classified as T type with the central core 
component shifted towards the trailing cone. The emission has been measured 
over a wide frequency range between 325 MHz and 10 GHz, with the core vanishing
at the highest frequencies. The relative intensity between the core and the 
cones show a power law dependence between 325 MHz and 5 GHz with 
$\Delta\alpha_{core/cone}\sim$ -0.5 (Fig. \ref{fig:spect2}, fourth row, right
column). \\

\emph{B2111+46} : The pulsar has a relatively wide profile that occupies more 
than 20 percent of the period and shows three distinct emission components with
T classification (see Fig. \ref{fig:avgprof}, right column). The pulsar 
emission is scattered at frequencies around 100 MHz. The power law dependence 
is seen for the frequency evolution of the relative intensity between 400 MHz 
and 5 GHz with $\Delta\alpha_{core/cone}\sim$ -0.8 (Fig. \ref{fig:spect3}, 
upper left panel). \\

\emph{B2210+29} : The pulsar has a T type profile and has many measurements 
between 100 MHz and 1.6 GHz. However, some of these profiles have lower 
detection sensitivities where the components cannot be measured. The relative 
intensity between the core and the cone has a power law dependence between 100 
MHz and 600 MHz with $\Delta\alpha_{core/cone}\sim$ -0.3 which signifies a 
relatively flatter behaviour (Fig. \ref{fig:spect3}, upper right panel), and 
shows further flattening at 1.4 GHz. \\

\emph{B2224+65} : The pulsar has a profile with two emission components and was
classified as T$_{1/2}$ with the leading component being the core and the 
trailing one the cone \citep{BMR15}. The profiles have been measured over 
multiple frequencies between 100 MHz and 1.6 GHz, but the cone vanishes around 
100 MHz. The detection sensitivities show large variations as a result of which
we used the peak intensities for the relative intensity estimates. The relative
intensity of the core and the cone shows a power law dependence between 325 MHz 
and 1.6 GHz with $\Delta\alpha_{core/cone}\sim$ -1.0 (see Fig. 
\ref{fig:spect3}, lower panel). \\

\section{Curvature Radiation}\label{sec:specmodl}

The radio emission in pulsars arises due to coherent curvature radiation from 
relativistic charge bunches, that are formed due to two stream instability 
developing in the outflowing plasma \citep{AM98,MGP00,GLM04}. The two stream 
instability can develop in overlapping clouds of plasma moving along the open 
magnetic field lines. The presence of an inner acceleration region (IAR), with 
unscreened electric field, above the polar cap is required to generate a 
non-stationary flow resulting in the overlapping clouds of plasma. The 
electron-positron pairs are formed due to magnetic pair creation from high 
energy photons in the IAR, which gets separated and accelerated into opposite 
directions by the electric field \citep{RS75,GMG03}. A sparking discharge 
ensues as a result of additional pair production from curvature radiation 
and/or inverse Compton scattering of high energy photons from the initial 
charges \citep{ML07,SMG15}. The potential difference along the IAR vanishes 
over timescales of hundreds of nanoseconds due to the excess charges which 
inhibits further pair formation. The positrons are accelerated away from the 
surface to relativistic energies, with Lorentz factors of $\gamma_p\sim10^6$, 
forming the primary particles. Once the IAR becomes sufficiently empty due to 
outflow of the excess charges, the electric field reappears starting the 
sparking process once again to generate the non-stationary plasma flow. 

The primary particles continue to radiate high energy photons as they move 
along the curved magnetic field lines beyond the IAR, resulting in a pair 
cascade that forms the secondary plasma comprising of both electron and 
positron streams. The primary particles outside the IAR has the co-rotation 
Goldreich-Julian density, $n_{GJ}$ \citep{GJ69}, while the secondary plasma has 
Lorentz factors of $\gamma_s\sim10^2$ and density $n_s$ = $\kappa n_{GJ}$, 
where $\kappa\geq10^4$ is the multiplicative factor \citep{S71} and $n_{GJ}$ 
can be obtained from : 
\begin{eqnarray}
n_{GJ} & = & -(\vec{\Omega}\cdot\vec{B})/2\pi c e \nonumber\\
       & = & ~~~ 5.6 \times 10^{5} \left(\frac{\dot{P}_{-15}}{P}\right)^{1/2} R_{50}^{-3} ~~~~ \rm{cm^{-3}}
\end{eqnarray}
Here $\vec{\Omega}$ is the angular velocity of stellar rotation, $\vec{B}$ is 
the magnetic field at a specific location, $P$ is the pulsar period, 
$\dot{P}_{-15} = \dot{P}/10^{-15}$ with $\dot{P}$ being the rate of period 
slowdown, $R_{50}$ = $r/50R_S$, where $r$ is the radial distance and $R_S=10$ 
km is the radius of the neutron star. The plasma instability develops in the 
secondary plasma, and its linear and nonlinear growth results in the formation 
of relativistic charge bunches around heights of 100-1000 km from the surface. 
As these charge bunches accelerate in the curved magnetic field lines they 
radiate curvature radiation which is seen as the radio emission.

The curvature of the magnetic field lines at any emission height increases from
the magnetic axis towards the edge of the open field line region. The centrally
located core component in the emission beam is expected to originate from the 
field lines close to the magnetic axis, while the inner and outer cones from 
field lines further away from the axis. In this section we explore the effect 
of the different magnetic field curvature on the spectra of the curvature 
radiation. Below we consider two different cases, the first corresponding to 
spectra of incoherent emission from a distribution of charged particles 
emitting curvature radiation and the second case deals with curvature radiation
from charged solitons. We are primarily concerned with the total spectra 
obtained from the radiation energy of these sources. However, quantities like 
the brightness temperature requires estimation of the radiation intensity 
\citep[see][]{RG19,GHW21}. The formation of charged solitons in the pulsar 
magnetosphere have been worked out in more detail in several works 
\citep{MGP00,LMM18,YZ18,RMM20} and is outside the scope of this paper. We use 
the results from these earlier works to simulate the variation in the spectral 
nature across the different profile components.

\subsection{Incoherent curvature radiation from distribution of charges}

\begin{figure}
\gridline{\fig{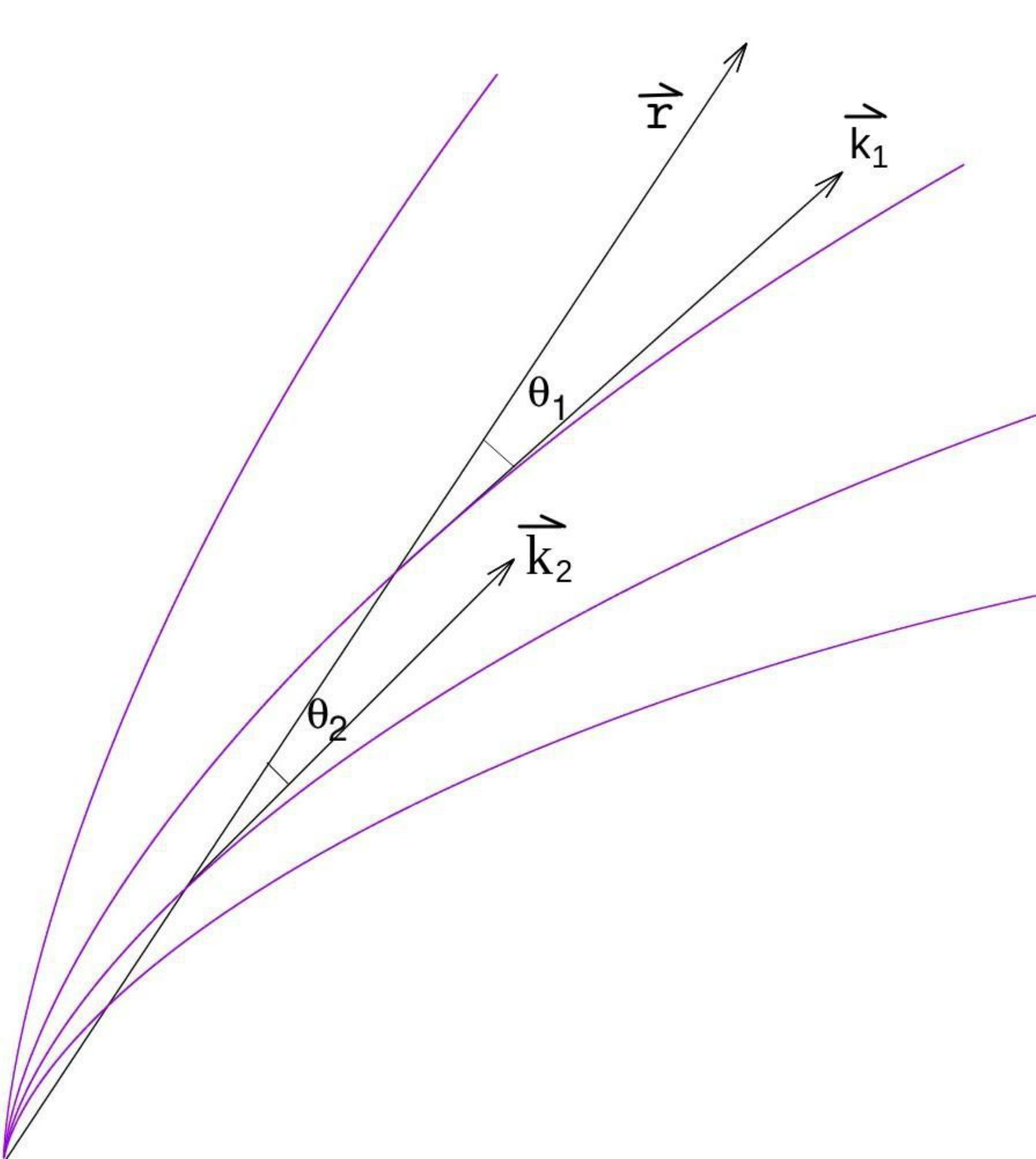}{0.38\textwidth}{}}
\caption{The figure shows a schematic representing the intersection of the line
of sight (LOS) along the radial direction ($\vec{r}$) with the magnetic field
lines. The wave vector of the radiation ($\vec{k}$) is along the tangent to the
magnetic field lines and the LOS makes different angles with the the wave
vector emitted from different heights, eg. $\theta_1$ and $\theta_2$ as shown
here. The curvature radiation around the characteristic frequency is 
highly beamed and can only be detected over a small angular radius $\theta < 
1/\gamma$, where $\gamma$ is the Lorentz factor of the relativistic particles. 
\label{fig:beam}}
\end{figure}

We first consider the case of the incoherent curvature radiation from a
distribution of charge particles with Lorentz factors equivalent to the 
secondary plasma, moving along the open dipolar magnetic field lines. This 
requires suitably adding the radiation energies from all particles visible 
within the line of sight for all frequencies. The emitted radiation energy 
($I$) at a given frequency ($\omega$) per unit solid angle ($\Omega$) by a 
relativistically charge particle moving along a curved trajectory can be 
obtained as \citep{J98} :
\begin{equation}
\frac{d^2I}{d\omega d\Omega} = I_0~\left(\frac{\omega}{\omega_c}\right)^2~(1 + \gamma_s^2\theta^2)^2~\left[K_{2/3}^2(\xi) + \left(\frac{\gamma_s^2\theta^2}{1 + \gamma_s^2\theta^2}\right)K_{1/3}^2(\xi)\right].
\end{equation}
where $I_0$ is the energy at the characteristic frequency ($\omega_c$) emitted 
along the tangential direction, $\gamma_s$ is the Lorentz factor of the 
secondary plasma particle, $\theta$ is the angle between the line of sight and 
the tangential direction of particle trajectory (see Fig. \ref{fig:beam}), 
$K_{1/3}$ and $K_{2/3}$ are modified Bessel functions, and
\begin{equation}
\omega_c = \frac{3}{2} \gamma_s^3 \left(\frac{c}{\rho_c}\right); ~~~~ \xi = \frac{\omega}{2\omega_c} \left(1 + \gamma_s^2\theta^2\right)^{3/2}
\end{equation}
with $\rho_c$ being the radius of curvature of the curved trajectory at the 
point of emission. The above spectrum peaks around $\omega\sim\omega_c$ and 
falls of sharply on either side.

In order to estimate the spectra for any particular line of sight, specified by
the angle $\theta$, the directed emission from all relevant charges needs to be
added for all frequencies. The radio emission is expected to arise between 
heights of $r=100$ km and $r=1000$ km. We consider the angular size of the 
core, inner cone and outer cone as defined below and average the spectra of all
line of sights within each component. These estimates can be compared with the 
observed spectra as shown in section \ref{sec:compspec}. We have used the 
following scheme to implement the measurement of spectra in each component. 

\begin{figure}
\gridline{\fig{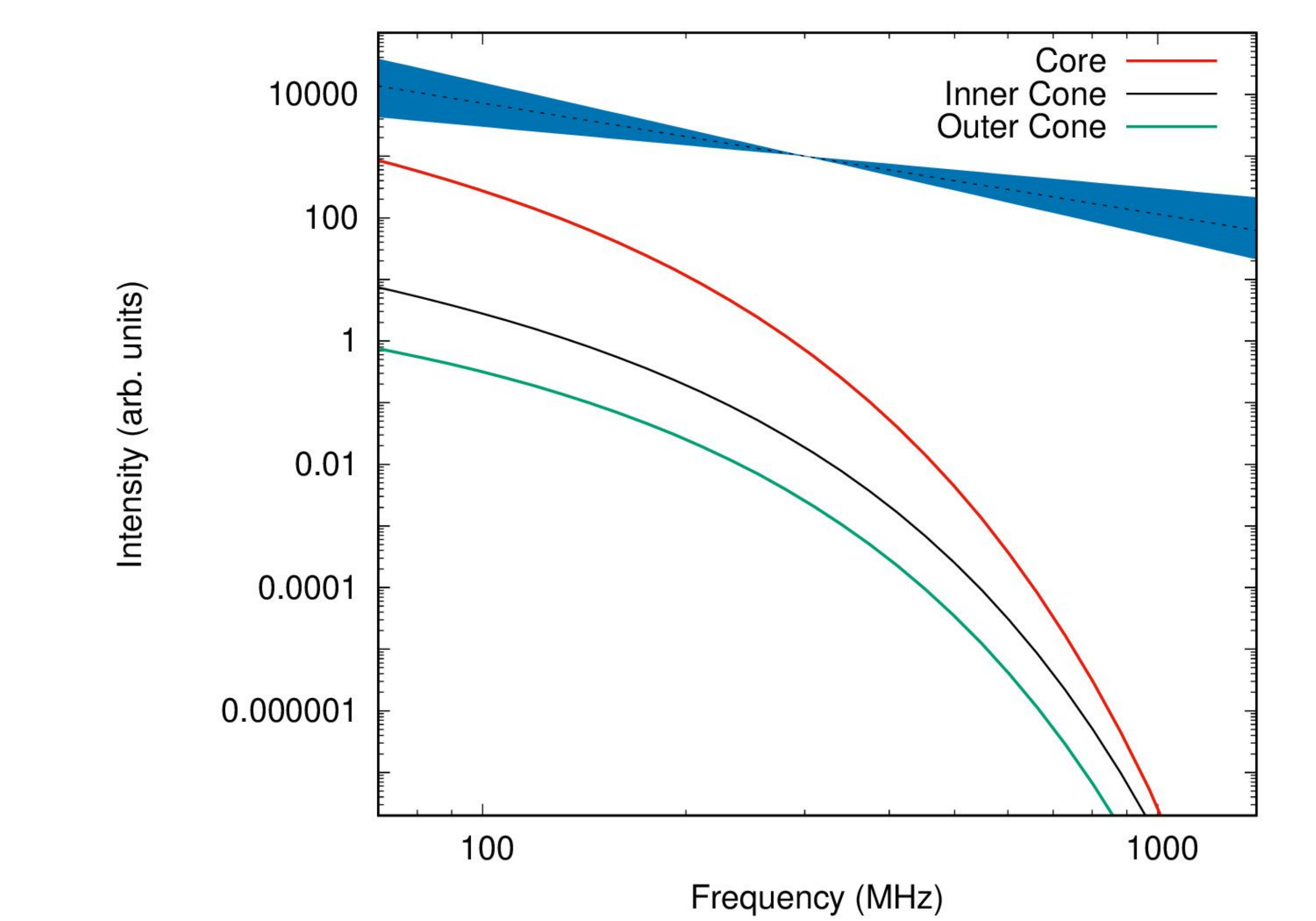}{0.48\textwidth}{}}
\caption{The figure shows the spectra of the different profile components from 
a distribution of relativistic particles emitting curvature radiation. The 
shaded region (blue) shows the expected emission spectra in the pulsar 
population with spectral index usually varying between -1.0 and -2.5 with a 
median value of -1.8 (dotted line).
\label{fig:snglspect}}
\end{figure}

\begin{itemize}
\item A power-law function, $n(\gamma) \propto \gamma^{a}$, is considered for 
the relativistic charge particle distribution. For these estimates $\gamma_s$ 
is confined between 50 and 300 with index $a=-0.3$. 

\item The emission is constrained to originate between heights of $10R_S < r < 
100R_S$. The radius of curvature at any given point is estimated in 
\citet[][see appendix]{BMM20b}.

\item At any given height the beam opening angle ($\theta_b$) is obtained as :
\begin{equation}
\sin{\theta_b} = \left(\frac{r}{R_{LC}}\right)^{1/2} ~~~ R_{LC} = Pc/2\pi.
\label{eq:maxopen}
\end{equation}
In order to estimate the spectra across the emission beam we consider the LOS
to cut centrally through the magnetic axis across the open field lines (which 
is indeed the case for M and T type pulsars mostly studied here). 

\item A relatively simple model of the emission beam is considered with core 
($\phi_{cr}$), inner cone ($\phi_{in}$) and outer cone ($\phi_{out}$) at
any given height defined as :
\begin{displaymath}
0 < \phi_{cr} < \theta_b/5, ~~~ \theta_b/5 < \phi_{in} < 3\theta_b/5, ~~~ 3\theta_b/5 < \phi_{out} < \theta_b.
\end{displaymath}
In addition the intensities are convolved with Gaussian functions centered 
around 0, $2\theta_b/5$ and $4\theta_b/5$ corresponding to the centers of the 
core, inner and outer cones, respectively, to emulate the observed profiles.
\end{itemize}

The radiation energies corresponding to all relevant particles within the 
confines of each component type and over the relevant emission heights are 
added up to form their respective spectra as shown in Fig. \ref{fig:snglspect}.
The simulated spectra shows a consistent evolution across the profile with the 
core component having a steeper spectra compared to the cones. This shows that 
the variations in the relative spectra between the core and the cones (see 
section \ref{sec:compspec}) can be naturally explained from curvature radiation
over narrow emission heights due to change in the radius of curvature from the 
axis towards the edge of the open field lines. However, as seen in Fig. 
\ref{fig:snglspect}, the spectra is much steeper than the observed spectra with
spectral index being around -4 to -6 between 100 MHz and 1 GHz compared to 
being around -2.0~in observed cases \citep{MKK00}. In order to obtain the 
measured spectra one requires $\gamma\sim10^3-10^4$ that are too high for the 
secondary plasma in pulsars. It should also be noted that for a distribution of
particles incoherent emission from individual particles are possible only when 
the separation between them is more than the wavelength of the emitted 
radiation. For longer wavelengths the particles will interact with each other 
resulting in absorption of radiation. This makes it impossible for the density 
of plasma in the pulsar magnetosphere to emit incoherent emission, where the 
average separation between particles $\propto n^{-1/3}$ is much smaller than 
radio wavelengths. Nonetheless, the purpose of this exercise is to demonstrate 
the effect of radius of curvature on the spectra of curvature radiation.

\subsection{Coherent Curvature radiation from Solitons}

The formation of charge separated solitons has been proposed by \citet[][MGP00
hereafter]{MGP00} as the mechanism for charge bunching in the pulsar 
magnetosphere. The two stream instability in the outflowing plasma clouds 
trigger strong Langmuir oscillations in the electron-positron plasma around 
heights of several hundred kilometers from the stellar surface \citep{AM98}. 
The amplitude of these waves are modulationally unstable and form packets of 
high density regions in the form of solitons. In the charge separated pulsar 
magnetosphere a natural difference arises between the energy distribution of 
electrons and positrons resulting in non-zero $\Delta\gamma = 
|\gamma_--\gamma_+|$, such that the relativistic mass difference between the 
two species produces inertial charge separation within the soliton envelope. 
These charge separated solitons emit coherent curvature radiation as they move 
along the open magnetic field lines. We use the details of soliton 
characteristics as described in MGP00 to simulate the spectra in the different 
profile components.

The typical longitudinal length scales of solitons along the magnetic field 
lines can be estimated as (see eq. 10~in MGP00) :
\begin{equation}
\Delta_s \sim 40 \gamma_2^{0.5} \kappa_4^{-0.5} R_{50}^{1.5} \Delta_d \chi^{-0.5} P^{0.25} \dot{P}_{-15}^{-0.25} ~~ \rm{cm,}
\end{equation}
where $\gamma_2=\gamma_s/100$, $\kappa_4=\kappa/10^4$, $\Delta_d\sim0.5$ is a 
dimensionless parameter and $\chi\sim0.1$ is related to the growth rate of the 
waves in the plasma. The typical length scales of solitons are around 10 to 100 
cm which ensures the emission of coherent radio emission ($\lambda > 
\Delta_s$). The spectrum of the total energy emitted by solitons can be 
estimated as :
\begin{equation}
\frac{d^2I}{d\omega d\Omega} = \frac{Q^2}{c} \omega_o F\left(\frac{\omega}{\omega_o},\theta\right) \left[1 - \cos\left(a\frac{\omega}{\omega_o}\right)\right]^2 
\end{equation}
Here $Q$ is the total charge in solitons, $\omega_o=\frac{3}{2}\gamma_o^3/\rho_c$, $a=\gamma_o^3(\Delta_s/\rho_c)$ and  
\begin{equation}
F\left(\frac{\omega}{\omega_o},\theta\right) = \left(\frac{\omega}{\omega_o}\right)^2~(1 + \gamma_o^2\theta^2)^2~\left[K_{2/3}^2(\xi) + \left(\frac{\gamma_o^2\theta^2}{1 + \gamma_o^2\theta^2}\right)K_{1/3}^2(\xi)\right], ~~~~ \xi = \frac{\omega}{2\omega_o} \left(1 + \gamma_o^2\theta^2\right)^{3/2}.
\end{equation}
In comparison with the single particle curvature radiation the soliton spectra 
has an extra term $\left[1 - \cos\left(a(\omega/\omega_o)\right)\right]^2$ 
which shifts the peak frequency by a factor of $\sim$4 from $\omega_o$ and 
makes the curve wider and more symmetrical (see Fig. 4~in MGP00). In addition 
the Lorentz factors of solitons $\gamma_o$ corresponds to the group velocities
of the wave packets and are usually greater than the Lorentz factors of the 
secondary plasma particles, $\gamma_o = y\gamma_s$, with $y\sim2$. As a result 
the characteristic frequencies of solitons $\omega_o$ are almost an order of 
magnitude greater than the single particle $\omega_c$. This further highlights 
the applicability of the soliton model as a viable mechanism for pulsar 
emission since the peak frequencies of soliton spectra are around 100-1000 MHz
which are between one and two orders of magnitude higher than the single 
particle case. Using basic pulsar parameters MGP00 estimated the spectra of the
coherent curvature radiation energy from each soliton as :
\begin{equation}
\frac{d^2I}{d\omega d\Omega} = A_1 \gamma_2 \kappa_4 R_{50}^{2.5} \chi^3 P^{-3/7} \dot{P}_{-15}^{-9/14} F\left(\frac{\omega}{\omega_o},\theta\right) \left[1 - \cos\left(a\frac{\omega}{\omega_o}\right)\right]^2 ~~\rm{ergs~Hz^{-1}~sr^{-1}},
\label{eq:solspec}
\end{equation}
and the total number of solitons contributing to the pulsar emission at any 
instant to be :
\begin{equation}
N_s \sim 10^5 \gamma_2^{-0.5} \kappa_4 R_{50}^{-0.5} \Delta_d^{-1} \chi^{0.5} P^{-1/4} \dot{P}_{-15}^{-1/4}.
\label{eq:solnum}
\end{equation}
The total energy of coherent curvature radiation emitted at any frequency can 
be obtained from $N_sdI/d\omega$. We are primarily interested in the relative 
difference of the spectra and the absolute values of the energy is contained 
in the constant term $A_1$. The soliton spectra obtained from 
eq.(\ref{eq:solspec}) and (\ref{eq:solnum}) is dependent on a number of 
parameters that are likely to have a range of values, e.g. the variability of 
soliton lengths, the Lorentz factors of solitons, the distribution of secondary
pair plasma, the optimal emission heights for the plasma instabilities to 
develop and their frequency dependence, the growth factor of the plasma waves, 
soliton charge densities, etc. It is unlikely that the full extent of how these
parameters vary can be estimated from observational constraints alone and would
require more detailed modelling of the non-linear plasma instabilities 
\citep[see for e.g.][]{LMM18,RMM20} which is outside the scope of this paper. 
It is expected that most of these parameters will vary as a function of distance
as well as across different field lines. However, our objective in this 
exercise is to determine if it is possible to obtain the observed variation in 
spectra seen across different components within the pulsar emission beam by 
using coherent curvature radiation from solitons. A similar process as 
explained for the incoherent case was followed to estimate the total radiation 
energy spectra across each component type, the core, the inner cone and the 
outer cone. We describe below the steps of estimating the spectra for 
different components using the above setup.

\begin{figure}
\gridline{\fig{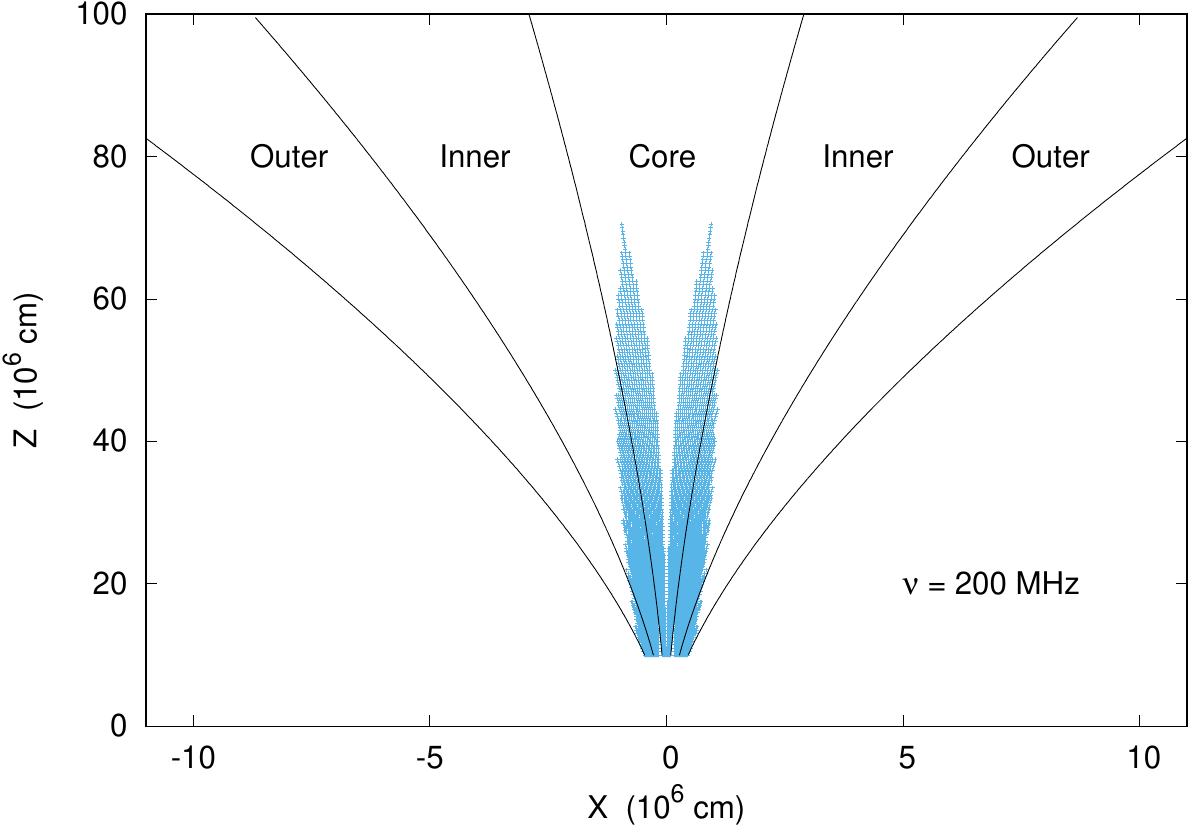}{0.48\textwidth}{(a)}}
\gridline{\fig{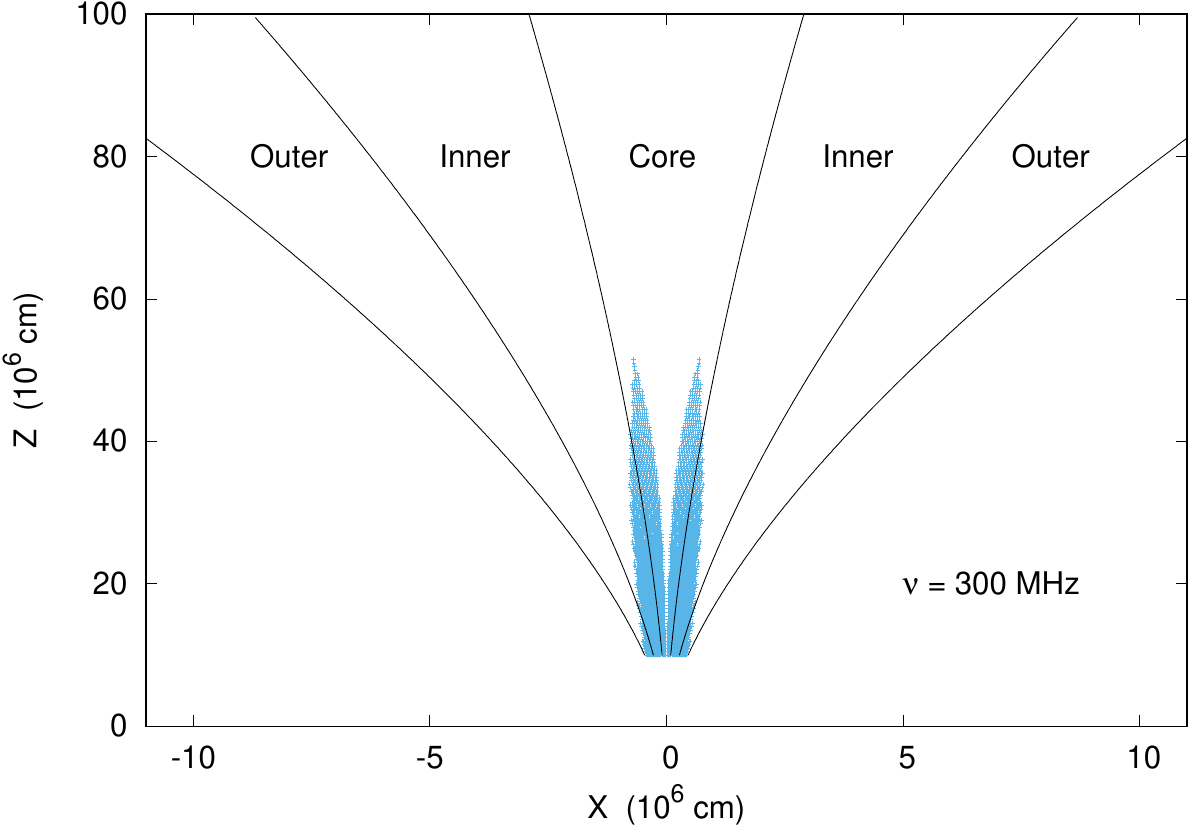}{0.48\textwidth}{(b)}}
\gridline{\fig{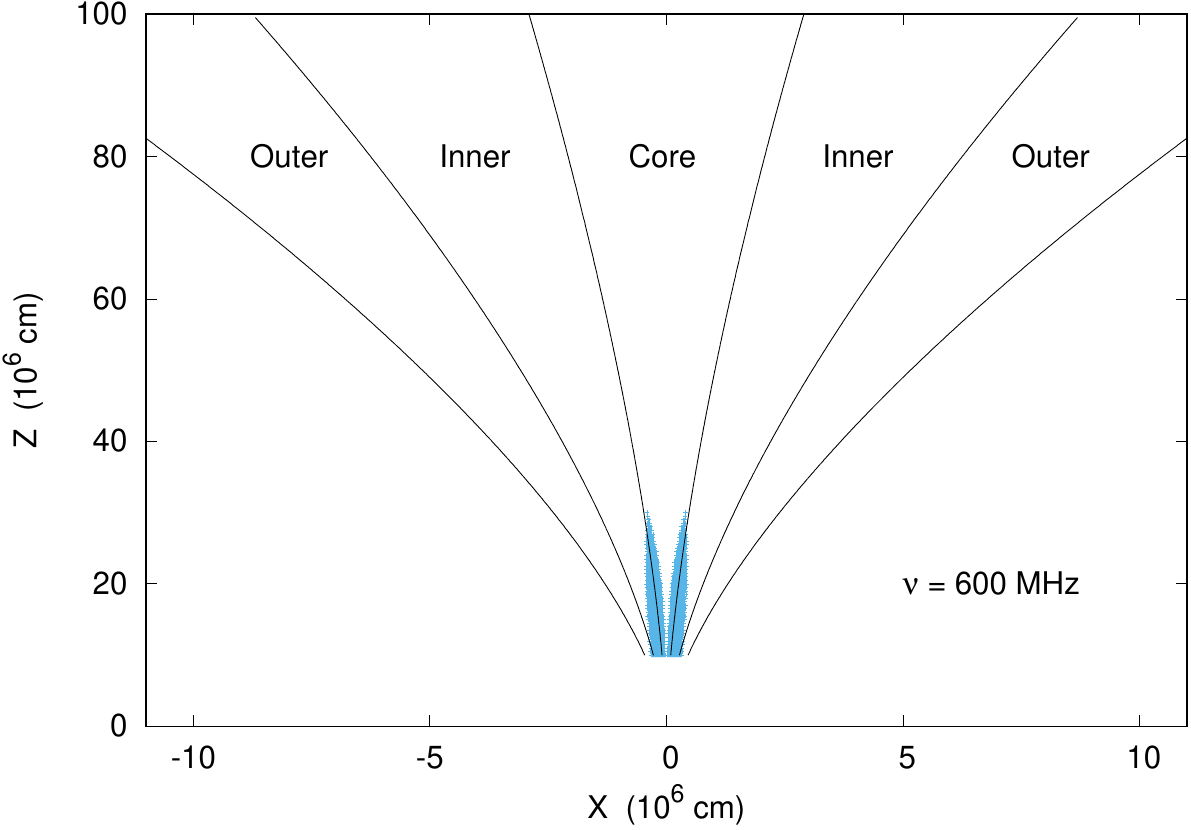}{0.48\textwidth}{(c)}}
\caption{The figure represents the projection of the open field line region for
an aligned rotator along the x-z plane where the shaded (blue) area corresponds
to the location of the simulated coherent curvature radio emission at (a) 200 
MHz, (b) 300 MHz and (c) 600 MHz. The lower limit for the emission height is 
fixed at 100 km (10$^7$ cm) and the boundaries of the regions corresponding to 
the core, inner cone and outer cone used in estimating the spectra are also 
shown. 
\label{fig:em_ht}}
\end{figure}

\begin{figure}
\gridline{\fig{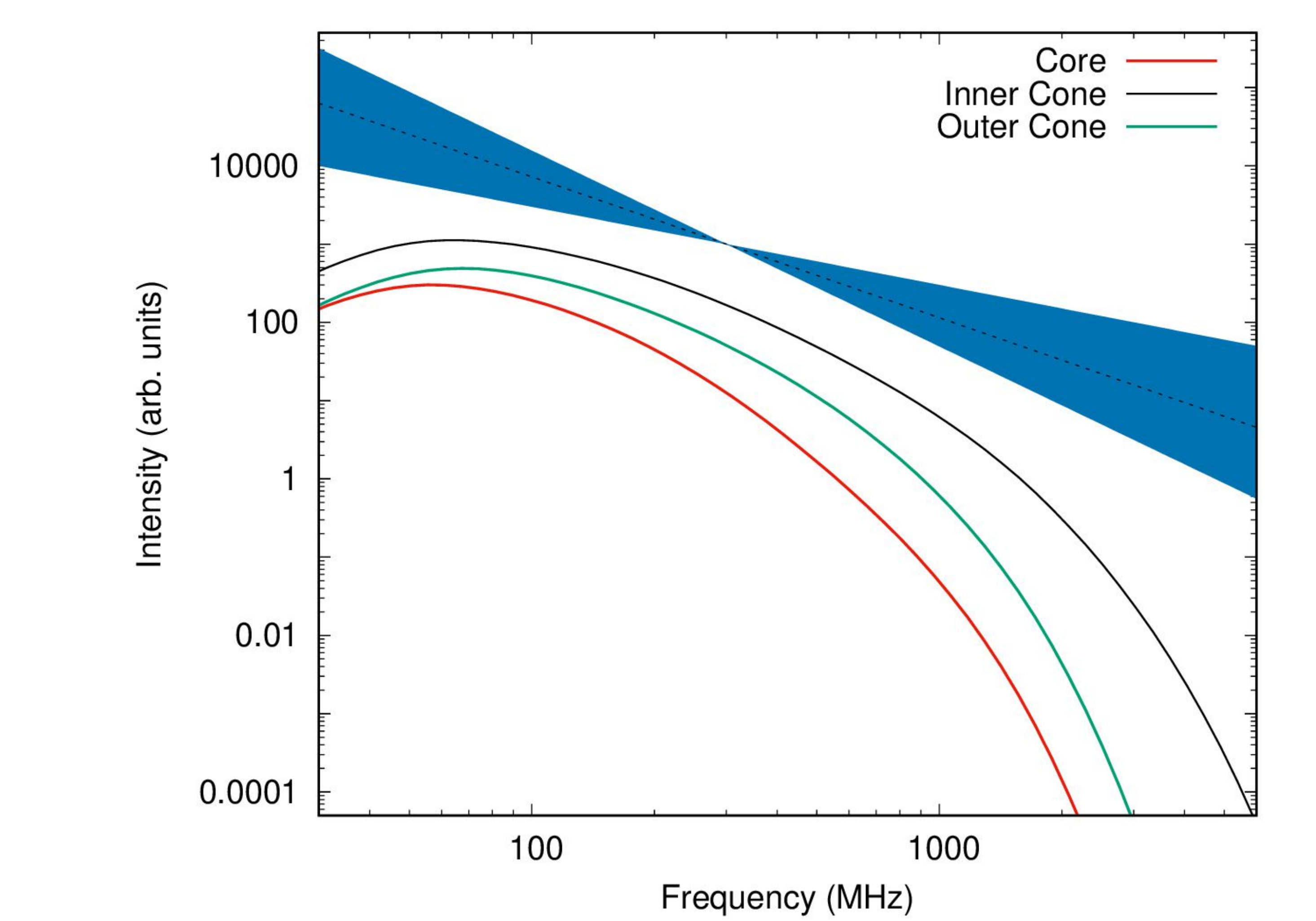}{0.48\textwidth}{(a)}
          \fig{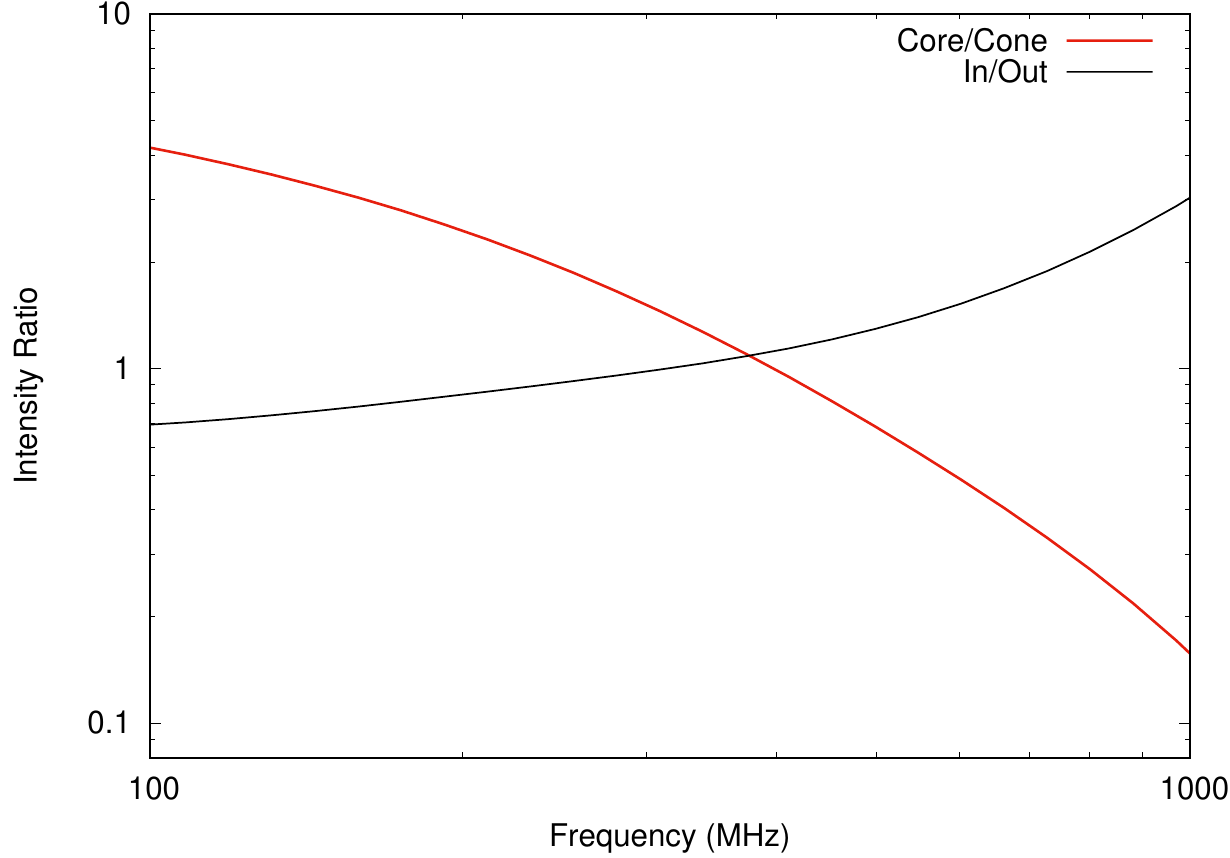}{0.48\textwidth}{(b)}}
\caption{The figure shows the variation in the spectra across the pulse window 
due to coherent curvature radiation from charged solitons. (a) The spectra of 
the core, inner cone and outer conal components. The shaded region (blue) shows
the expected spectral nature of emission in the pulsar population with spectral
index usually varying between -1.0 and -2.5 with a median value of -1.8 (dotted
line). (b) the frequency evolution of the relative intensities in the core and 
cone as well as the inner and outer cones. The ratios of the intensities have 
been scaled to fit within the window.
\label{fig:solspect}}
\end{figure}

\begin{itemize}
\item The emission is once gain constrained to originate between heights of 
$10R_S < r < 100R_S$.

\item The emission beam model is identical to the one used for the single 
particle curvature radiation described earlier, comprising of a central core 
between $0 < \phi_{cr} < \theta_b/5$, an inner cone between $\theta_b/5 < 
\phi_{in} < 3\theta_b/5$ and an outer cone with $3\theta_b/5 < \phi_{out} <
\theta_b$, where $\theta_b$ is defined in eq.(\ref{eq:maxopen}). We consider a 
central LOS traverse across the beam and approximate the components to have 
Gaussian shapes.

\item For most parameters we use a fixed set of values as prescribed in MGP00.
The only variable parameter is the Lorentz factors of secondary plasma $\gamma$
which are allowed to have different set of values around $\gamma\sim100$ in 
order to vary the shape of spectra across the components. We once again 
consider a power law distribution for the Lorentz factors with index $a=-0.3$. 
The range of $\gamma$ used for the core is between 50 and 170, the inner cone 
between 50 and 280, and the outer cone with variations from 5 to 130.
\end{itemize}

We are primarily interested in the behaviour of the spectra within the 
frequency range 100 MHz and 1 GHz where the relative component spectra are well 
constrained. The region from where a particular frequency in emitted depends on
$\omega_o$ which is a function of the radius of curvature and $\gamma_o$. The 
radius of curvature further depends on the emission height as well as the 
distance from the magnetic axis. For any given $\gamma_o$ the emission along 
any field line is directed towards the observer when the angle between the 
radial vector ($\vec{r}$) and the tangent to the field line at the point of 
emission is less than $1/\gamma_o$ (see Fig. \ref{fig:beam}). This introduces 
an additional constraint on the emission between the inner magnetic field 
lines, closer to the magnetic axis and the outer field lines further away 
particularly those for the outer cones where the curvature is higher resulting 
in larger angles. Fig. \ref{fig:em_ht} shows the area of the open field line 
region where the radio emission arises from at different frequencies of 200 
MHz, 300 MHz and 600 MHz for our selected setup of emission parameters. We 
divided the open field line region between emission heights of 100 to 1000 km 
into small grids and separately estimated the spectra in each grid. For any 
given frequency if significant emission was seen at a grid, i.e. more that 1 
percent of power of the grid with maximum energy for that frequency, it was 
marked with a blue point and shown in the figure. At the lower frequencies the
emission can originate from much higher $r$, while the emission is restricted 
to much lower $r$ in the outer cones across all frequencies. Fig. 
\ref{fig:solspect} shows the estimated spectra for the core, inner cone and 
outer cones. The core emission is steeper than the cones due to variation in 
the curvature which is also seen for the single particle case. The outer cones 
also have steeper spectra than the inner cones as a result of the beaming 
effect such that emission from certain field lines at higher frequencies are 
not directed towards the observer making the spectra steeper. The spectral 
index for the inner cone between 100 MHz and 1 GHz is -2.0 (\ref{fig:solspect},
left panel), which is consistent with observations. In addition 
$\Delta\alpha_{core/cone}\sim$-1.2 and $\Delta\alpha_{in/out}\sim$+0.6 
(\ref{fig:solspect}, right panel) which matches the observational results in 
section \ref{sec:compspec}. It should be noted that the physical parameters 
used in these simulations of spectra are demonstrative in nature and do not 
represent unique solutions regarding the physical conditions in the pulsar 
magnetosphere.

\section{Conclusion}
The relative spectral index between the core and the conal emission has been
estimated by \cite{BMM21} in 21 pulsars over a relatively narrow frequency 
range between 325 MHz and 610 MHz. They found the core emission to be 
relatively steeper than the cones with mean relative spectral index of 
$\Delta\alpha_{core/cone}\sim-0.7$. The outer cone also showed a steeper 
spectra than the inner cone such that $\Delta\alpha_{in/out}\sim+0.4$. In this 
work we have used average profile measurements from archival observations to 
expand the above studies in a larger sample of 53 pulsars with much wider 
frequency coverage between 100 MHz and 10 GHz. In all cases we found the core 
component to have a steeper spectra compared to the core, although the 
differences in spectra show a relatively uniform distribution between $-0.2$ 
and $-2.0$. In 9 pulsars we were able to measure the the spectral difference 
between the inner and the outer cones, with the inner cone being less steep 
than the outer cone with relative spectral index between $+0.1$ and $+0.8$. In 
a few cases the differences in spectra show flattening at higher frequencies 
above 1 GHz. However, in almost all such cases we found that these were a 
result of some of the components merging together at higher frequencies thereby
affecting relative intensity estimates. Our analysis clearly highlight the 
evolution of the pulsar spectra within the emission window in the entire pulsar
population as no clear correlation is seen between the relative spectral index 
and different pulsar parameters.

The radio emission in pulsars is expected to originate within a narrow range of
heights, a few hundred kilometers from the stellar surface, due to coherent 
curvature radiation. This requires the formation of charged bunches in the form
of charge separated solitons due to non-linear instabilities in the outflowing 
plasma. A natural steepening of the spectra is expected due to increase of 
curvature of the magnetic field lines from the axis towards the edge of the 
open field lines. Further, the relativistic beaming effect restricts the 
emission at the outermost field lines from being directed towards the observer.
This mostly affects the higher frequencies which are supposed to be emitted by 
particles with higher Lorentz factors. As a result additional steepening of the
spectra is expected for the outermost field lines compared to relatively inner 
ones. The above mechanism can explain the variations in spectra seen in the 
pulsar emission window if one considers the core to arise from field lines 
close to the magnetic axis, the inner cones from field lines further away from 
the core, while the outer cones are furthest away and located near the edge of 
the open field line region. 

The plasma populating the open field line region originates due to sparking 
discharges in an inner acceleration region (IAR) above the polar cap. The 
sparking process is thermally regulated due to discharge of ions from the 
heated polar cap surface forming a partially screened gap \citep[PSG,][]{GMG03,
SMG15}. Differences in the expected potential drop along the IAR can arise from
the center of the IAR towards the polar cap edge due to the distribution of 
sparks. The center is populated by sparks from all sides and is likely to have 
a larger potential drop, while the thermal regulation requires the presence of 
a spark always near the rim of the polar cap which likely has reduced potential
difference along the gap \citep{GS00,BMM20b}. As a result the distribution 
function of the outflowing plasma generated from the sparks can show variations
between the central regions and the outer field lines. Our studies show that 
such variations are likely to affect the relative differences in spectra across 
the profile. However, the emission mechanism is dependent on a number of 
parameters which are not well constrained. More detailed modelling of the 
nonlinear plasma processes are required to narrow the range of these parameters
before such association between the variations in spectra and the plasma 
distribution function can be studied in more detail.

\section*{Acknowledgments}
We thank the referee for comments that helped to improve the paper. DM 
acknowledges the support of the Department of Atomic Energy, Government of 
India, under project no. 12-R\&D-TFR-5.02-0700. DM acknowledges funding from 
the grant ``Indo-French Centre for the Promotion of Advanced Research - 
CEFIPRA" grant IFC/F5904-B/2018. This work was supported by the grant 
2020/37/B/ST9/02215 of the National Science Centre, Poland.

\bibliography{CompSpect}{}
\bibliographystyle{aasjournal}

\startlongtable
\begin{longrotatetable}
\begin{deluxetable}{ccccccccccccc}
\tablenum{3}
\tablecaption{Component Spectral difference\label{tab:spect}}
\tablewidth{0pt}
\tablehead{  & \colhead{PSR} & \colhead{Type} & \colhead{Freq} & \colhead{$S_{core}$/$S_{cone}$} & \colhead{$\Delta\alpha$} & \colhead{$S_{core}$/$S_{in}$} & \colhead{$\Delta\alpha$} & \colhead{$S_{core}$/$S_{out}$} & \colhead{$\Delta\alpha$} & \colhead{$S_{in}$/$S_{out}$} & \colhead{$\Delta\alpha$} &  Ref. \\
  &   &   &  (MHz)  &   &   &   &   &   &   &   &  }

\startdata
 1 & B0203-40 & T$_{1/2}$ & 325 & 11.9$\pm$0.9 & -0.53$\pm$0.07 & --- & --- & --- & --- & --- & --- & [1] \\
  &   &   & 1400 & 5.5$\pm$0.4 &   & --- & --- & --- & --- & --- & --- & [2] \\
  &   &   &   &   &   &   &   &   &   &   &  &  \\
 2 & B0329+54 & T & 140 & 4.076$\pm$0.007 & -0.72$\pm$0.06 & --- & --- & --- & --- & --- & --- & [3] \\
  &   &   & 408 & 6.830$\pm$0.006 & (408--2250) & --- & --- & --- & --- & --- & --- & [4] \\
  &   &   & 610 & 5.555$\pm$0.002 &   & --- & --- & --- & --- & --- & --- & [4] \\
  &   &   & 925 & 3.926$\pm$0.004 &   & --- & --- & --- & --- & --- & --- & [4] \\
  &   &   & 1410 & 2.339$\pm$0.002 &   & --- & --- & --- & --- & --- & --- & [5] \\
  &   &   & 1642 & 2.707$\pm$0.005 &   & --- & --- & --- & --- & --- & --- & [4] \\
  &   &   & 2250 & 2.190$\pm$0.004 &   & --- & --- & --- & --- & --- & --- & [6] \\
  &   &   & 4850 & 3.265$\pm$0.004 &   & --- & --- & --- & --- & --- & --- & [5] \\
  &   &   & 8500 & 3.5$\pm$0.1 &   & --- & --- & --- & --- & --- & --- & [6] \\
  &   &   & 10550 & 1.72$\pm$0.02 &   & --- & --- & --- & --- & --- & --- & [8] \\
  &   &   &   &   &   &   &   &   &   &   &  &  \\
 3 & B0450+55 & T & 325 & 3.27$\pm$0.01 & -0.41$\pm$0.02 & --- & --- & --- & --- & --- & --- & [7] \\
  &   &   & 408 & 2.91$\pm$0.06 & (325--1642) & --- & --- & --- & --- & --- & --- & [4] \\
  &   &   & 610 & 2.39$\pm$0.02 &   & --- & --- & --- & --- & --- & --- & [4] \\
  &   &   & 910 & 2.08$\pm$0.03 &   & --- & --- & --- & --- & --- & --- & [4] \\
  &   &   & 1400 & 1.809$\pm$0.006 &   & --- & --- & --- & --- & --- & --- & [8] \\
  &   &   & 1642 & 1.71$\pm$0.05 &   & --- & --- & --- & --- & --- & --- & [4] \\
  &   &   & 4850 & 1.41$\pm$0.02 &   & --- & --- & --- & --- & --- & --- & [5] \\
  &   &   &   &   &   &   &   &   &   &   &  &  \\
 4 & B0621-04 & M & 410 & 0.9$\pm$0.1 & -0.72$\pm$0.06 & --- & --- & --- & --- & --- & --- & [4] \\
  &   &   & 610 & 0.60$\pm$0.03 &   & --- & --- & --- & --- & --- & --- & [4] \\
  &   &   & 1408 & 0.33$\pm$0.01 &   & --- & --- & --- & --- & --- & --- & [4] \\
  &   &   &   &   &   &   &   &   &   &   &  &  \\
 5 & B0626+24 & T$_{1/2}$ & 170 & 4.2$\pm$0.1 & -1.03$\pm$0.05 & --- & --- & --- & --- & --- & --- & [9] \\
  &   &   & 325 & 4.79$\pm$0.09 & (325--610) & --- & --- & --- & --- & --- & --- & [1] \\
  &   &   & 610 & 2.51$\pm$0.06 &   & --- & --- & --- & --- & --- & --- & [1] \\
  &   &   & 4850 & 0.99$\pm$0.03 &   & --- & --- & --- & --- & --- & --- & [10] \\
  &   &   &   &   &   &   &   &   &   &   &  &  \\
 6 & B0844-35 & $_c$Q & 325 & --- & --- & --- & --- & --- & --- & 1.60$\pm$0.02 & 0.61$\pm$0.03 & [1] \\
  &   &   & 610 & --- & --- & --- & --- & --- & --- & 2.34$\pm$0.04 &  & [1] \\
  &   &   &   &   &   &   &   &   &   &   &  &  \\
 7 & B0919+06 & T & 135 & 4.84$\pm$0.03 & -0.60$\pm$0.01 & --- & --- & --- & --- & --- & --- & [3] \\
  &   &   & 325 & 2.90$\pm$0.01 &   & --- & --- & --- & --- & --- & --- & [1] \\
  &   &   & 610 & 1.94$\pm$0.01 &   & --- & --- & --- & --- & --- & --- & [1] \\
  &   &   &   &   &   &   &   &   &   &   &  &  \\
 8 & B0940-55 & T & 1400 & 3.60$\pm$0.02 & -1.40$\pm$0.04 & --- & --- & --- & --- & --- & --- & [2] \\
  &   &   & 3100 & 1.19$\pm$0.04 &   & --- & --- & --- & --- & --- & --- & [11] \\
  &   &   &   &   &   &   &   &   &   &   &  &  \\
 9 & J1034-3224 & $_c$Q & 325 & --- & --- & --- & --- & --- & --- & 0.617$\pm$0.003 & 0.78$\pm$0.05 & [1] \\
  &  &  & 408 & --- & --- & --- & --- & --- & --- & 0.766$\pm$0.008 & (325--610) & [12] \\
  &  &  & 610 & --- & --- & --- & --- & --- & --- & 1.024$\pm$0.005 &  & [1] \\
  &  &  & 1400 & --- & --- & --- & --- & --- & --- & 0.798$\pm$0.007 &  & [2] \\
  &  &  &  &  &  &  &  &  &  &  &  &  \\
10 & B1046-58 & $_c$T & 1400 & --- & --- & --- & --- & --- & --- & 1.545$\pm$0.004 & 0.22$\pm$0.02 & [2] \\
  &  &  & 8356 & --- & --- & --- & --- & --- & --- & 2.28$\pm$0.03 &  & [13] \\
  &  &  &  &  &  &  &  &  &  &  &  &  \\
11 & J1141-3322 & T & 436 & 9.9$\pm$1.0 & -1.61$\pm$0.14 & --- & --- & --- & --- & --- & --- & [14] \\
  &  &  & 1400 & 1.51$\pm$0.06 &  & --- & --- & --- & --- & --- & --- & [2] \\
  &  &  &  &  &  &  &  &  &  &  &  &  \\
12 & B1154-62 & T & 1400 & 5.29$\pm$0.03 & -1.50$\pm$0.06 & --- & --- & --- & --- & --- & --- & [2] \\
  &  &  & 3100 & 1.61$\pm$0.07 &  & --- & --- & --- & --- & --- & --- & [11] \\
  &  &  &  &  &  &  &  &  &  &  &  &  \\
13 & B1237+25 & M & 120 & 1.044$\pm$0.003 & -0.32$\pm$0.04 & 1.413$\pm$0.006 & -0.46$\pm$0.06 & 0.950$\pm$0.003 & -0.29$\pm$0.04 & 0.673$\pm$0.002 & 0.15$\pm$0.03 & [9] \\
  &  &  & 180 & 0.966$\pm$0.007 & (120--610) & 1.23$\pm$0.01 & (120--610) & 0.880$\pm$0.006 & (120--610) & 0.715$\pm$0.006 & (120--610) & [9] \\
  &  &  & 325 & 0.727$\pm$0.002 &  & 0.822$\pm$0.002 &  & 0.680$\pm$0.001 &  & 0.827$\pm$0.001 &  & [1] \\
  &  &  & 610 & 0.648$\pm$0.004 &  & 0.716$\pm$0.005 &  & 0.612$\pm$0.004 &  & 0.855$\pm$0.003 &  & [1] \\
  &  &  & 1400 & 0.709$\pm$0.002 &  & 0.637$\pm$0.002 &  & 0.758$\pm$0.002 &  & 1.190$\pm$0.002 &  & [8] \\
  &  &  &  &  &  &  &  &  &  &  &  &  \\
14 & B1323-58 & T$_{1/2}$ & 1400 & 6.20$\pm$0.08 & -1.64$\pm$0.08 & --- & --- & --- & --- & --- & --- & [11] \\
  &  &  & 3100 & 1.63$\pm$0.04 &  & --- & --- & --- & --- & --- & --- & [11] \\
  &  &  & 8356 & 0.43$\pm$0.08 &  & --- & --- & --- & --- & --- & --- & [13] \\
  &  &  &  &  &  &  &  &  &  &  &  &  \\
15 & B1325-49 & M & 325 & 0.102$\pm$0.001 & -0.28$\pm$0.04 & 0.367$\pm$0.005 & -0.36$\pm$0.04 & 0.142$\pm$ 0.001 & -0.24$\pm$0.04 & 0.387$\pm$0.003 & 0.12$\pm$0.02 & [1] \\
  &  &  & 610 & 0.086$\pm$0.002 &  & 0.292$\pm$0.007 &  & 0.122$\pm$0.002 &  & 0.417$\pm$0.005 &  & [1] \\
  &  &  &  &  &  &  &  &  &  &  &  &  \\
16 & B1353-62 & T & 1400 & 1.012$\pm$0.003 & -1.86$\pm$0.02 & --- & --- & --- & --- & --- & --- & [2] \\
  &  &  & 3100 & 0.231$\pm$0.005 &  & --- & --- & --- & --- & --- & --- & [11] \\
  &  &  &  &  &  &  &  &  &  &  &  &  \\
17 & B1508+55 & T & 140 & 8.791$\pm$0.004 & -0.78$\pm$0.08 & --- & --- & --- & --- & --- & --- & [9] \\
  &  &  & 180 & 6.895$\pm$0.003 &  & --- & --- & --- & --- & --- & --- & [9] \\
  &  &  & 325 & 4.591$\pm$0.003 &  & --- & --- & --- & --- & --- & --- & [15] \\
  &  &  &  &  &  &  &  &  &  &  &  &  \\
18 & B1541+09 & T & 140 & 26.5$\pm$0.5 & -1.22$\pm$0.06 & --- & --- & --- & --- & --- & --- & [9] \\
  &  &  & 180 & 18.8$\pm$0.6 &  & --- & --- & --- & --- & --- & --- & [9] \\
  &  &  & 610 & 4.66$\pm$0.06 &  & --- & --- & --- & --- & --- & --- & [4] \\
  &  &  & 800 & 2.89$\pm$0.06 &  & --- & --- & --- & --- & --- & --- & [16] \\
  &  &  & 1418 & 2.139$\pm$0.008 &  & --- & --- & --- & --- & --- & --- & [17] \\
  &  &  &  &  &  &  &  &  &  &  &  &  \\
19 & J1557-4258 & T & 610 & 7.4$\pm$0.3 & -0.72$\pm$0.05 & --- & --- & --- & --- & --- & --- & [1] \\
  &  &  & 1400 & 4.11$\pm$0.06 &  & --- & --- & --- & --- & --- & --- & [2] \\
  &  &  &  &  &  &  &  &  &  &  &  &  \\
20 & B1556-44 & T & 325 & 11.8$\pm$0.8 & -0.64$\pm$0.13 & --- & --- & --- & --- & --- & --- & [1] \\
  &  &  & 1400 & 5.42$\pm$0.03 &  & --- & --- & --- & --- & --- & --- & [2] \\
  &  &  & 1560 & 3.6$\pm$0.1 &  & --- & --- & --- & --- & --- & --- & [18] \\
  &  &  &  &  &  &  &  &  &  &  &  &  \\
21 & B1600-49 & T & 610 & 9.0$\pm$0.2 & -0.34$\pm$0.04 & --- & --- & --- & --- & --- & --- & [1] \\
  &  &  & 1400 & 6.81$\pm$0.06 &  & --- & --- & --- & --- & --- & --- & [2] \\
  &  &  &  &  &  &  &  &  &  &  &  &  \\
22 & J1625-4048 & T & 436 & 0.88$\pm$0.01 & -0.83$\pm$0.23 & --- & --- & --- & --- & --- & --- & [12] \\
  &  &  & 610 & 0.576$\pm$0.002 &  & --- & --- & --- & --- & --- & --- & [1] \\
  &  &  & 1400 & 0.36$\pm$0.02 &  & --- & --- & --- & --- & --- & --- & [12] \\
  &  &  &  &  &  &  &  &  &  &  &  &  \\
23 & B1642-03 & T & 610 & 52.1$\pm$0.4 & -1.39$\pm$0.05 & --- & --- & --- & --- & --- & --- & [1] \\
  &  &  & 1400 & 14.7$\pm$0.3 &  & --- & --- & --- & --- & --- & --- & [2] \\
  &  &  & 4850 & 3.07$\pm$0.07 &  & --- & --- & --- & --- & --- & --- & [5] \\
  &  &  & 10550 & 1.05$\pm$0.07 &  & --- & --- & --- & --- & --- & --- & [8] \\
  &  &  &  &  &  &  &  &  &  &  &  &  \\
24 & B1700-32 & T & 325 & 0.601$\pm$0.001 & -0.477$\pm$0.005 & --- & --- & --- & --- & --- & --- & [1] \\
  &  &  & 610 & 0.445$\pm$0.001 &  & --- & --- & --- & --- & --- & --- & [1] \\
  &  &  &  &  &  &  &  &  &  &  &  &  \\
25 & B1732-07 & T & 325 & 8.55$\pm$0.07 & -0.69$\pm$0.05 & --- & --- & --- & --- & --- & --- & [1] \\
  &  &  & 610 & 5.8$\pm$0.1 &  & --- & --- & --- & --- & --- & --- & [1] \\
  &  &  & 1400 & 3.00$\pm$0.08 &  & --- & --- & --- & --- & --- & --- & [2] \\
  &  &  &  &  &  &  &  &  &  &  &  &  \\
26 & B1737+13 & M & 325 & 1.887$\pm$0.009 & -0.78$\pm$0.09 & 6.17$\pm$0.05 & -1.12$\pm$0.13 & 2.78$\pm$0.01 & -0.78$\pm$0.11 & 0.450$\pm$0.005 & 0.27$\pm$0.07 & [1] \\
  &  &  & 408 & 1.57$\pm$0.06 &  & 4.5$\pm$0.3 &  & 2.4$\pm$0.1 &  & 0.54$\pm$0.04 &  & [4] \\
  &  &  & 610 & 0.977$\pm$0.003 &  & 2.77$\pm$0.01 &  & 1.551$\pm$0.006 &  & 0.560$\pm$0.003 &  & [1] \\
  &  &  & 910 & 0.89$\pm$0.04 &  & 2.2$\pm$0.1 &  & 1.36$\pm$0.06 &  & 0.62$\pm$0.04 &  & [1] \\
  &  &  & 1418 & 0.645$\pm$0.001 &  & 1.780$\pm$0.006 &  & 1.013$\pm$0.003 &  & 0.569$\pm$0.002 &  & [17] \\
  &  &  &  &  &  &  &  &  &  &  &  &  \\
27 & B1738-08 & $_c$Q & 325 & --- & --- & --- & --- & --- & --- & 1.206$\pm$0.009 & 0.36$\pm$0.02 & [1] \\
  &  &  & 610 & --- & --- & --- & --- & --- & --- & 1.52$\pm$0.01 &  & [1] \\
  &  &  &  &  &  &  &  &  &  &  &  &  \\
28 & B1758-29 & T & 325 & 1.65$\pm$0.02 & -0.89$\pm$0.05 & --- & --- & --- & --- & --- & --- & [1] \\
  &  &  & 610 & 0.91$\pm$0.02 &  & --- & --- & --- & --- & --- & --- & [1] \\
  &  &  & 1400 & 0.48$\pm$0.01 &  & --- & --- & --- & --- & --- & --- & [2] \\
  &  &  &  &  &  &  &  &  &  &  &  &  \\
29 & B1804-08 & T & 610 & 1.67$\pm$0.01 & -0.87$\pm$0.12 & --- & --- & --- & --- & --- & --- & [1] \\
  &  &  & 925 & 1.32$\pm$0.02 &  & --- & --- & --- & --- & --- & --- & [4] \\
  &  &  & 1400 & 0.757$\pm$0.001 &  & --- & --- & --- & --- & --- & --- & [2] \\
  &  &  & 3100 & 0.399$\pm$0.008 &  & --- & --- & --- & --- & --- & --- & [11] \\
  &  &  &  &  &  &  &  &  &  &  &  &  \\
30 & B1821+05 & T & 325 & 5.64$\pm$0.04 & -1.55$\pm$0.10 & --- & --- & --- & --- & --- & --- & [1] \\
  &  &  & 408 & 4.2$\pm$0.2 & (325--925) & --- & --- & --- & --- & --- & --- & [4] \\
  &  &  & 610 & 2.05$\pm$0.04 &  & --- & --- & --- & --- & --- & --- & [1] \\
  &  &  & 925 & 1.11$\pm$0.03 &  & --- & --- & --- & --- & --- & --- & [4] \\
  &  &  & 1408 & 0.98$\pm$0.01 &  & --- & --- & --- & --- & --- & --- & [4] \\
  &  &  & 1642 & 0.87$\pm$0.04 &  & --- & --- & --- & --- & --- & --- & [4] \\
  &  &  & 4850 & 0.450$\pm$0.006 &  & --- & --- & --- & --- & --- & --- & [10] \\
  &  &  &  &  &  &  &  &  &  &  &  &  \\
31 & B1826-17 & T & 925 & 1.12$\pm$0.02 & -1.98$\pm$0.04 & --- & --- & --- & --- & --- & --- & [4] \\
  &  &  & 1408 & 0.490$\pm$0.002 & --- & --- & --- & --- & --- & --- & --- & [4] \\
  &  &  &  &  &  &  &  &  &  &  &  &  \\
32 & B1831-03 & T & 925 & 6.6$\pm$0.5 & -0.74$\pm$0.25 & --- & --- & --- & --- & --- & --- & [4] \\
  &  &  & 1400 & 4.8$\pm$0.3 &  & --- & --- & --- & --- & --- & --- & [8] \\
  &  &  &  &  &  &  &  &  &  &  &  &  \\
33 & B1831-04 & M & 325 & 2.60$\pm$0.02 & -0.80$\pm$0.02 & 3.19$\pm$0.04 & -1.22$\pm$0.02 & 2.40$\pm$0.02 & -0.62$\pm$0.02 & 0.75$\pm$0.01 & 0.60$\pm$0.03 & [1] \\
  &  &  & 610 & 1.57$\pm$0.01 &  & 1.48$\pm$0.01 &  & 1.63$\pm$0.01 &  & 1.10$\pm$0.01 &  & [1] \\
  &  &  &  &  &  &  &  &  &  &  &  &  \\
34 & B1839+09 & T$_{1/2}$ & 130 & 3.6$\pm$0.2 & -0.56$\pm$0.07 & --- & --- & --- & --- & --- & --- & [9] \\
  &  &  & 170 & 3.0$\pm$0.1 & (130--325) & --- & --- & --- & --- & --- & --- & [9] \\
  &  &  & 325 & 2.17$\pm$0.07 &  & --- & --- & --- & --- & --- & --- & [1] \\
  &  &  & 4850 & 0.89$\pm$0.06 &  & --- & --- & --- & --- & --- & --- & [10] \\
  &  &  &  &  &  &  &  &  &  &  &  &  \\
35 & B1857-26 & M & 325 & 1.016$\pm$0.003 & -1.00$\pm$0.07 & 3.09$\pm$0.01 & -1.44$\pm$0.09 & 1.566$\pm$0.005 & -0.92$\pm$0.16 & 0.507$\pm$0.002 & 0.35$\pm$0.20 & [1] \\
  &  &  & 410 & 0.786$\pm$0.007 &  & 2.23$\pm$0.03 &  & 1.15$\pm$0.01 &  & 0.564$\pm$0.007 &  & [4] \\
  &  &  & 610 & 0.522$\pm$0.002 &  & 1.154$\pm$0.005 &  & 0.967$\pm$0.005 &  & 0.786$\pm$0.004 &  & [1] \\
  &  &  & 925 & 0.320$\pm$0.003 &  & 0.766$\pm$0.009 &  & 0.528$\pm$0.006 &  & 0.690$\pm$0.008 &  & [4] \\
  &  &  & 1400 & 0.283$\pm$0.001 &  & --- & --- & --- & --- & --- & --- & [2] \\
  &  &  &  &  &  &  &  &  &  &  &  &  \\
36 & B1859+03 & T$_{1/2}$ & 925 & 9.8$\pm$1.0 & -1.69$\pm$0.10 & --- & --- & --- & --- & --- & --- & [4] \\
  &  &  & 1418 & 4.6$\pm$0.2 &  & --- & --- & --- & --- & --- & --- & [17] \\
  &  &  & 1642 & 3.9$\pm$0.3 &  & --- & --- & --- & --- & --- & --- & [4] \\
  &  &  &  &  &  &  &  &  &  &  &  &  \\
37 & B1907-03 & T & 610 & 7.3$\pm$0.5 & -1.16$\pm$0.08 & --- & --- & --- & --- & --- & --- & [4] \\
  &  &  & 1420 & 2.75$\pm$0.06 &  & --- & --- & --- & --- & --- & --- & [8] \\
  &  &  &  &  &  &  &  &  &  &  &  &  \\
38 & B1907+00 & T & 610 & 11.6$\pm$0.6 & -1.38$\pm$0.10 & --- & --- & --- & --- & --- & --- & [4] \\
  &  &  & 1418 & 3.6$\pm$0.2 &  & --- & --- & --- & --- & --- & --- & [17] \\
  &  &  &  &  &  &  &  &  &  &  &  &  \\
39 & B1907+10 & T$_{1/2}$ & 610 & 13.7$\pm$1.4 & -1.52$\pm$0.15 & --- & --- & --- & --- & --- & --- & [1] \\
  &  &  & 1400 & 3.9$\pm$0.3 &  & --- & --- & --- & --- & --- & --- & [2] \\
  &  &  &  &  &  &  &  &  &  &  &  &  \\
40 & B1911+13 & T & 606 & 5.1$\pm$0.4 & -0.32$\pm$0.14 & --- & --- & --- & --- & --- & --- & [4] \\
  &  &  & 1418 & 3.9$\pm$0.3 &  & --- & --- & --- & --- & --- & --- & [17] \\
  &  &  &  &  &  &  &  &  &  &  &  &  \\
41 & B1914+09 & T$_{1/2}$ & 325 & 2.93$\pm$0.07 & -0.78$\pm$0.11 & --- & --- & --- & --- & --- & --- & [1] \\
  &  &  & 410 & 3.0$\pm$0.2 &  & --- & --- & --- & --- & --- & --- & [4] \\
  &  &  & 610 & 1.84$\pm$0.06 &  & --- & --- & --- & --- & --- & --- & [1] \\
  &  &  & 925 & 1.3$\pm$0.1 &  & --- & --- & --- & --- & --- & --- & [4] \\
  &  &  & 1400 & 1.11$\pm$0.04 &  & --- & --- & --- & --- & --- & --- & [2] \\
  &  &  & 1642 & 0.78$\pm$0.07 &  & --- & --- & --- & --- & --- & --- & [4] \\
  &  &  &  &  &  &  &  &  &  &  &  &  \\
42 & B1917+00 & T & 325 & 1.86$\pm$0.01 & -0.57$\pm$0.07 & --- & --- & --- & --- & --- & --- & [1] \\
  &  &  & 408 & 1.69$\pm$0.09 &  & --- & --- & --- & --- & --- & --- & [4] \\
  &  &  & 610 & 1.29$\pm$0.02 &  & --- & --- & --- & --- & --- & --- & [1] \\
  &  &  & 1418 & 1.38$\pm$0.05 &  & --- & --- & --- & --- & --- & --- & [17] \\
  &  &  & 1642 & 1.46$\pm$0.12 &  & --- & --- & --- & --- & --- & --- & [4] \\
  &  &  &  &  &  &  &  &  &  &  &  &  \\
43 & B1918+26 & T$_{1/2}$ & 170 & 2.07$\pm$0.08 & -0.28$\pm$0.03 & --- & --- & --- & --- & --- & --- & [9] \\
  &  &  & 1418 & 1.14$\pm$0.07 &  & --- & --- & --- & --- & --- & --- & [17] \\
  &  &  &  &  &  &  &  &  &  &  &  &  \\
44 & B1920+21 & T$_{1/2}$ & 610 & 6.3$\pm$0.2 & -1.03$\pm$0.21 & --- & --- & --- & --- & --- & --- & [4] \\
  &  &  & 925 & 3.4$\pm$0.8 &  & --- & --- & --- & --- & --- & --- & [4] \\
  &  &  & 1418 & 2.4$\pm$0.3 &  & --- & --- & --- & --- & --- & --- & [17] \\
  &  &  & 1642 & 2.7$\pm$0.3 &  & --- & --- & --- & --- & --- & --- & [4] \\
  &  &  &  &  &  &  &  &  &  &  &  &  \\
45 & B1929+10 & T & 120 & 2.31$\pm$0.06 & -0.178$\pm$0.004 & --- & --- & --- & --- & --- & --- & [9] \\
  &  &  & 140 & 2.27$\pm$0.05 & (120--4750) & --- & --- & --- & --- & --- & --- & [9] \\
  &  &  & 180 & 2.15$\pm$0.04 &  & --- & --- & --- & --- & --- & --- & [9] \\
  &  &  & 325 & 1.920$\pm$0.004 &  & --- & --- & --- & --- & --- & --- & [1] \\
  &  &  & 410 & 1.83$\pm$0.03 &  & --- & --- & --- & --- & --- & --- & [19] \\
  &  &  & 610 & 1.68$\pm$0.01 &  & --- & --- & --- & --- & --- & --- & [19] \\
  &  &  & 925 & 1.60$\pm$0.01 &  & --- & --- & --- & --- & --- & --- & [4] \\
  &  &  & 1710 & 1.480$\pm$0.006 &  & --- & --- & --- & --- & --- & --- & [5] \\
  &  &  & 4750 & 1.21$\pm$0.02 &  & --- & --- & --- & --- & --- & --- & [10] \\
  &  &  & 10550 & 1.23$\pm$0.04 &  & --- & --- & --- & --- & --- & --- & [10] \\
  &  &  &  &  &  &  &  &  &  &  &  &  \\
46 & B1946+35 & T & 925 & 8.1$\pm$0.3 & -1.53$\pm$0.08 & --- & --- & --- & --- & --- & --- & [4] \\
  &  &  & 1418 & 4.5$\pm$0.1 &  & --- & --- & --- & --- & --- & --- & [17] \\
  &  &  & 1642 & 3.4$\pm$0.1 &  & --- & --- & --- & --- & --- & --- & [4] \\
  &  &  & 4850 & 0.60$\pm$0.08 &  & --- & --- & --- & --- & --- & --- & [5] \\
  &  &  &  &  &  &  &  &  &  &  &  &  \\
47 & B1952+29 & T & 610 & 1.51$\pm$0.07 & -1.91$\pm$0.06 & --- & --- & --- & --- & --- & --- & [4] \\
  &  &  & 1418 & 0.30$\pm$0.01 &  & --- & --- & --- & --- & --- & --- & [17] \\
  &  &  &  &  &  &  &  &  &  &  &  &  \\
48 & B2002+31 & T & 610 & 16.7$\pm$1.0 & -1.49$\pm$0.11 & --- & --- & --- & --- & --- & --- & [4] \\
  &  &  & 1418 & 5.15$\pm$0.09 &  & --- & --- & --- & --- & --- & --- & [17] \\
  &  &  & 1642 & 3.4$\pm$0.2 &  & --- & --- & --- & --- & --- & --- & [4] \\
  &  &  &  &  &  &  &  &  &  &  &  &  \\
49 & B2045-16 & T & 325 & 1.069$\pm$0.001 & -0.53$\pm$0.06 & --- & --- & --- & --- & --- & --- & [1] \\
  &  &  & 408 & 1.124$\pm$0.009 &  & --- & --- & --- & --- & --- & --- & [4] \\
  &  &  & 610 & 0.822$\pm$0.001 &  & --- & --- & --- & --- & --- & --- & [1] \\
  &  &  & 925 & 0.70$\pm$0.01 &  & --- & --- & --- & --- & --- & --- & [4] \\
  &  &  & 1400 & 0.514$\pm$0.001 &  & --- & --- & --- & --- & --- & --- & [2] \\
  &  &  & 1642 & 0.524$\pm$0.002 &  & --- & --- & --- & --- & --- & --- & [4] \\
  &  &  & 4850 & 0.207$\pm$0.007 &  & --- & --- & --- & --- & --- & --- & [5] \\
  &  &  &  &  &  &  &  &  &  &  &  &  \\
50 & B2111+46 & T & 408 & 3.54$\pm$0.04 & -0.83$\pm$0.03 & --- & --- & --- & --- & --- & --- & [4] \\
  &  &  & 610 & 2.205$\pm$0.008 &  & --- & --- & --- & --- & --- & --- & [4] \\
  &  &  & 800 & 1.57$\pm$0.01 &  & --- & --- & --- & --- & --- & --- & [16] \\
  &  &  & 925 & 1.66$\pm$0.02 &  & --- & --- & --- & --- & --- & --- & [4] \\
  &  &  & 1408 & 1.128$\pm$0.003 &  & --- & --- & --- & --- & --- & --- & [4] \\
  &  &  & 1642 & 1.03$\pm$0.01 &  & --- & --- & --- & --- & --- & --- & [4] \\
  &  &  & 4850 & 0.396$\pm$0.002 &  & --- & --- & --- & --- & --- & --- & [10] \\
  &  &  &  &  &  &  &  &  &  &  &  &  \\
51 & B2210+29 & T & 130 & 0.79$\pm$0.03 & -0.31$\pm$0.04 & --- & --- & --- & --- & --- & --- & [9] \\
  &  &  & 170 & 0.69$\pm$0.02 & (130--610) & --- & --- & --- & --- & --- & --- & [9] \\
  &  &  & 610 & 0.48$\pm$0.01 &  & --- & --- & --- & --- & --- & --- & [4] \\
  &  &  & 1418 & 0.466$\pm$0.006 &  & --- & --- & --- & --- & --- & --- & [17] \\
  &  &  &  &  &  &  &  &  &  &  &  &  \\
52 & B2224+65 & T$_{1/2}$ & 325 & 4.9$\pm$0.4 & -1.01$\pm$0.11 & --- & --- & --- & --- & --- & --- & [7] \\
  &  &  & 400 & 2.2$\pm$0.3 &  & --- & --- & --- & --- & --- & --- & [16] \\
  &  &  & 610 & 1.65$\pm$0.06 &  & --- & --- & --- & --- & --- & --- & [4] \\
  &  &  & 800 & 1.22$\pm$0.06 &  & --- & --- & --- & --- & --- & --- & [16] \\
  &  &  & 925 & 0.9$\pm$0.1 &  & --- & --- & --- & --- & --- & --- & [4] \\
  &  &  & 1408 & 0.71$\pm$0.03 &  & --- & --- & --- & --- & --- & --- & [4] \\
  &  &  & 1642 & 0.75$\pm$0.06 &  & --- & --- & --- & --- & --- & --- & [4] \\
  &  &  &  &  &  &  &  &  &  &  &  &  \\
53 & B2327-20 & T & 325 & 1.686$\pm$0.003 & -0.581$\pm$0.008 & --- & --- & --- & --- & --- & --- & [1] \\
  &  &  & 610 & 1.170$\pm$0.005 &  & --- & --- & --- & --- & --- & --- & [1] \\
  &  &  &  &  &  &  &  &  &  &  &  &  \\
\enddata
\tablecomments{ [1]-\cite{MBM16}; [2]-\cite{JK18}; [3]-\cite{PHS16}; 
[4]-\cite{GL98}; [5]-\cite{vHX97}; [6]-\cite{KXJ97}; [7]-\cite{MR11}; 
[8]-\cite{SGG95}; [9]-\cite{BKK16}; [10]-\cite{KKW98}; [11]-\cite{KJ06}; 
[12]-\cite{DSB98}; [13]-\cite{JKW06}; [14]-\cite{LYL95}; [15]-\cite{BMM20a};
[16]-\cite{ANT94}; [17]-\cite{WCL99}; [18]-\cite{WML93}; [19]-\cite{STC99}}
\end{deluxetable}
\end{longrotatetable}

\end{document}